\documentclass[a4paper,fleqn,usenatbib]{mnras}
\usepackage{newtxtext,newtxmath}

\usepackage[T1]{fontenc}
\usepackage{ae,aecompl}
\usepackage{comment}

\usepackage{graphicx}	
\usepackage{amsmath}	
\usepackage{epstopdf}
\usepackage{lineno}
\usepackage{xspace}
\usepackage{lineno}
\usepackage{multirow} 
\usepackage{longtable}
\usepackage{supertabular,booktabs}

\title[Flares in blazars]{Catching profound optical flares in blazars}

\author[Bhatta et al.]{
Gopal Bhatta$^{1}$\thanks{E-mail: gopal.bhatta@ifj.edu.pl},\ \
Staszek Zola$^{2}$,\ \
M. Drozdz$^{3}$,\ \
Daniel Reichart$^{4}$,\ \
Joshua Haislip$^{4}$,\ \
Vladimir Kouprianov$^{4}$,\ \
\newauthor Katsura Matsumoto$^{5}$,\ \
Eda Sonbas$^{6,7}$,\ \
D. Caton$^{8}$,\ \
Urszula Pajdosz-\'Smierciak$^{2}$,\ \
A. Simon$^{9}$,\ \
J. Provencal$^{10,11}$,\ \
\newauthor  Dariusz G\'ora$^{1}$,\ \
and Grzegorz Stachowski$^{2}$,\ \
\\
$^{1}$Institute of Nuclear Physics Polish Academy of Sciences, PL-31342 Krak\'ow, Poland\\
$^{2}$Astronomical Observatory of the Jagiellonian University,
 ul. Orla 171, 30-244 Krak\'ow, Poland\\
 $^{3}$Mt. Suhora Observatory, Pedagogical University,
 ul. Podchorazych 2, 30-084 Krak\'ow, Poland\\
  $^{4}$Dept. of Physics and Astronomy, University of North Carolina at Chapel Hill,
 Chapel Hill, NC 27599, USA\\
 $^{5}$Astronomical Institute, Osaka Kyoiku University,
4-698 Asahigaoka, Kashiwara, Osaka 582-8582, Japan\\
$^{6}$University of Adiyaman, Department of Physics,
02040 Adiyaman, Turkey\\
$^{7}$Astrophysics Application and Research Center, Adiyaman University, Adiyaman 02040, Turkey\\
$^{8}$Dark Sky Observatory, Dept. of Physics and Astronomy, Appalachian State University,
 Boone, NC 28608, USA\\
 $^{9}$Astronomy and Space Physics Department, Taras Shevshenko National University of Kyiv,
 Volodymyrska str. 60, 01033 Kyiv, Ukraine\\
 $^{10}$University of Delaware, Department of Physics and Astronomy Newark,
 DE 19716, USA\\
  $^{11}$Delaware Asteroseismic Research Center, Mt. Cuba Observatory,
 Greenville, DE 19807, USA\\
}

\date{Accepted XXX. Received YYY; in original form ZZZ}

\pubyear{2015}

\begin{document}
\label{firstpage}
\maketitle

\begin{abstract}
Flaring episodes in blazars represent one of the most violent processes observed in extra-galactic objects. Studies of such events shed light on the 
energetics of the physical processes occurring in the innermost regions of blazars, which cannot otherwise be resolved by any current instruments. In this work, we present some of the largest and most rapid flares captured in the optical band in the blazars 3C 279, OJ 49, S4 0954+658, TXS 1156+295 and PG 1553+113. The source flux was observed to increase by nearly ten times within a timescale of a few weeks. We applied several methods of time series analysis and symmetry analysis. Moreover, we also performed searches for periodicity in the light curves of 3C 279, OJ 49 and PG 1553+113 using the Lomb-Scargle method and found plausible indications of quasi-periodic oscillations (QPOs). In particular, the 33- and 22-day periods found in 3C 279, i.e. a 3:2 ratio, are intriguing. These violent events might originate from magnetohydrodynamical instabilities near the base of the jets, triggered by processes modulated by 
the magnetic field of the accretion disc. We present a qualitative treatment as the possible explanation for the observed large amplitude flux changes in both the source-intrinsic and source-extrinsic scenarios.

\end{abstract}



\section{Introduction }
\label{sec:intro}
Blazars are a sub-class of radio-loud active galactic nuclei (AGN) featuring relativistic jets which are closely aligned to the line of sight \citep{Urry1995}. The blazar continuum emission is non-thermal in nature, and it is Doppler boosted and highly variable over a wide range of spatial and temporal frequencies. Blazars consist of two kinds of sources: flat-spectrum radio quasars (FSRQ) and BL Lacertae (BL Lac) objects. Of the two types of sources, FSRQs show broad emission lines, while BL Lacs exhibit either weak emission lines or their absence over the continuum. Nonetheless, the objects are visible in the TeV energy range and constitute the dominant population of discrete gamma-ray sources in the sky. Historically, FSRQs are considered to be more luminous than  BL Lacs, however new reports suggest that some of the BL Lacs can be more luminous than FSRQs  \cite[see][]{Sheng2022}. The broadband non-thermal spectrum of blazars, which extends from radio to the highest energy $\gamma$-rays such as TeV emission, exhibits two distinct low- and high-energy components which respectively peak between radio and soft X-rays and between X-rays and $\gamma$-rays. The low-energy component is well explained in terms of synchrotron emission by relativistic plasma in the magnetized jets. However, various models, mainly leptonic or hadronic scenarios, have been put forward to explain the origin of high-energy emission. According to the leptonic scenario, ultra-relativistic electrons up-scatter low-energy seed photons into X-rays and $\gamma$-rays via the inverse Compton mechanism. Further, in the synchrotron self-Compton (SSC; e.g. \citealt{Marscher1985,Maraschi1992}) model, the synchrotron photons produced by the electrons constitute a radiation field which is inverse-Compton scattered by the co-spatial leptons. In the external Compton (EC) models, AGN components such as the accretion disc, broad-line region and dusty torus \citep[see][]{Ghisellini1996,Sikora1994,Dermer1992} could contribute the low-energy seed photons required for the inverse-Compton scattering. Conversely, the hadronic models suggest that the protons may be accelerated to very high energies, which then produce the high-energy spectral component via direct proton-synchrotron and/or photon-initiated cascades \citep{Mucke2003,Aharonian2000,Mannheim1993}. Although both leptonic and hadronic models can account for the origin of high-energy emission from blazars, the hadronic models require a stronger magnetic field to radiatively cool more massive protons. Similarly, the proton-synchrotron models also require the acceleration of ultra high energy cosmic rays (UHECRs) in the blazar jets, which trigger photo-pionic interactions resulting in gamma-ray emission along with some secondary particles. In such interactions, neutrino production in the jets becomes a natural outcome. Indeed, the IceCube \citep{IceCube2018Sci} experiment detected high energy neutrinos potentially associated with gamma-ray flaring in the blazar TXS 0506+056.

Blazars are characterized by multi-wavelength (MWL) flux variability over diverse timescales \citep[e.g. see][]{Bhatta2018a,bhatta16b, Bhatta2018c, Bhatta2020, Bhatta2021}. Although the statistical variability properties can be largely represented by a single power-law spectral density (see optical; \citealt[][]{Nilsson2018}; and $\gamma$-ray; \citealt{Bhatta2020}), blazars often display complex variability patterns, such as red-noise like variability superimposed by occasional sharp rises in the flux. In majority of the cases, such rises in flux possess a well-defined shape and last for a definite duration and therefore can be identified as distinct flaring events, which can either be observed simultaneously in MWL observations \citep[e.g.][]{Acciari2020,Abeysekara2018,Balokovic2016,Hayashida2015,Aleksic2015}, or only in a specific wavelength band, often termed as ``orphan'' flares, \citep[e.g. see][for optical orphan flares]{Chatterjee2013}. Some of the dominant flaring events, which show a large rise in the flux, typically last from a few weeks to months. These events suggest the presence of extreme physical conditions prevalent around the central engine as well as in the jets, which may drive the most efficient particle acceleration and cooling processes. In the gamma-ray domain, blazar flaring episodes are often accompanied by a large swing in the rotation of the plane of optical polarization \citep{Blinov2018}, consistent with violent collision between the relativistic shock waves and stationary structures such as Mach disc. Moreover, such events are also seen to be accompanied by ejection of radio knots seen in Very Long Baseline Array images \citep{Park2019,Marscher2010}. Studies of blazar radio jets seem to indicate that, in most cases, flaring events in blazars can be linked with disturbances propagating along the jet which lead to the ejection of radio knots showing apparent superluminal motion\citep{Jorstad2016,Kellermann2004,Jorstad2001}. In the case of weakly magnetized jets, the shock waves propagating along the jet can energize the particles to induce the flaring events \citep{Lind1985}. On the other hand, if the jets are highly magnetized, magnetic reconnection might play a dominant role in particle acceleration\citep{Giannios2013,Nalewajko2011} before energy dissipation. Also the re-collimation of the shock waves can lead to the formation of rapid flares \citep{Bromberg2009}

\begin{table*}
 \caption{General information about the sample of blazar targets \label{table:1} }
 \centering
 \begin{tabular}{l|l|l|l|c}
 \hline
 Source name & Source class &R.A. (J2000) & Dec. (J2000) & Redshift (z) \\
 \hline
 OJ 49 &BL Lac, LSP & $08^h31^m48.88^s$ & $+04^d29^m39.086^s$ & 0.17386\\
		S4 0954+658 & BL Lac, LSP& $09^h58^m47.2^s$ & $+65^d33^m55^s$ & 0.368 \\ 
 TXS 1156+295 &FSRQ, LSP & $11^h59^m032.07^s$ & $+29^d14^m42.0^s$ & 0.729 \\
 3C 279 &FSRQ, LSP& $12^h56^m11.1665^s$ & $-05^d47^m21.523^s$ & 0.536 \\
 PG 1553+113 &BL Lac, HSP& $15^h55^m43.044^s$ & $+11^d11^m24.365^s$ & 0.36 \\ 
 \hline
 \end{tabular}
 \end{table*}

In this work, we present the results of our analysis of flaring observations of five blazars obtained by our group in the course of long-term optical monitoring of AGN. In Section \ref{sec:2}, the observations of the source sample and the relevant optical data processing are described. A brief description of each of the sample sources is presented in Section \ref{sec:3}. In Section \ref{sec:4}, several analytical approaches using various methods including fractional variability, flux distribution, PSD and QPOs are introduced, and the results of the analyses of the light curves are presented. The results along with their possible implications are discussed in Section \ref{sec:5}, and we summarize our conclusions in Section \ref{sec:6}.

\section{Observations and data processing }
 \label{sec:2}
 Long term monitoring of a sample of quasars was primarily carried out using small telescopes operated by the Skynet Robotic Telescope Network \citep{Skynet}. Additional data were collected with the 60~cm telescope located at Adiyaman University Astrophysics Application and Research Center, Turkey, the 50~cm telescope at the Osaka University Observatory in Japan, and with two telescopes in Poland: a 60~cm at the Mt. Suhora Observatory and a 50~cm at the Jagiellonian University Observatory in Krakow. The wide band R filter (Bessell prescription) was most often used for the monitoring. Longer runs were also carried out, mostly at the Krakow and Mt. Suhora sites.

The data taken by Skynet consists of several scientific images of a target taken each night, subsequently reduced for bias, dark and flatfield by the network pipeline. Other sites provided raw images accompanied by calibration frames. We performed reduction of raw images with the standard procedure: calibration for bias, dark and ﬂatﬁeld (usually taken on the sky) with the IRAF package, 
while extraction of magnitudes was done using aperture photometry with the CMunipack program, 
which implements the DAOPHOT algorithm.
As a result, differential magnitudes were derived with comparison stars for each 
object chosen so as to be visible in the field of view of all telescopes. Their constancy 
was verified with check stars which were similarly chosen.
 The sample of sources along with their classes, positions and red-shifts are presented in Table \ref{table:1}, which also lists the blazar source classification based on synchrotron peak frequency, that is, high synchrotron peaked blazars (HSP; $\nu_\mathrm{peak}^S > 10^{15} $ Hz),  intermediate synchrotron peaked blazars (ISP; $10^{14} < \nu_\mathrm{peak}^S < 10^{15}$ Hz), and low synchrotron peaked blazars (LSP; $\nu_\mathrm{peak}^S < 10^{14}$Hz) \citep{{2010ApJ...716...30A}}.  Similarly, for a given source, the total observation duration, the number of observations and the mean magnitude are listed in the 2nd, 3rd and 4th columns, respectively, of Table \ref{table:2}. 
 
\begin{figure*}
\begin{center} 
{\includegraphics[width=0.48\textwidth,angle=0]{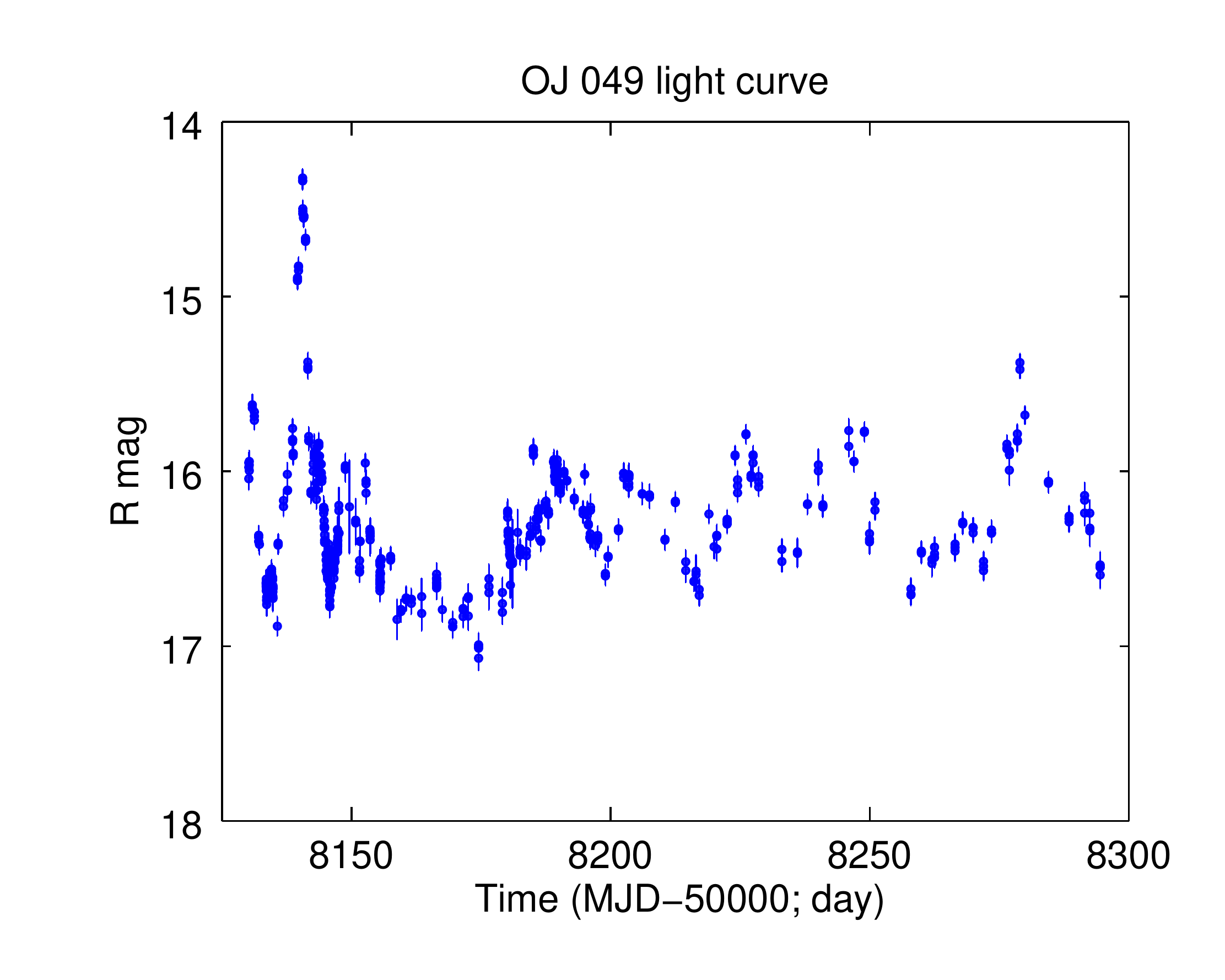}} 
\hspace{-0.1cm}
{\includegraphics[width=0.48\textwidth,angle=0]{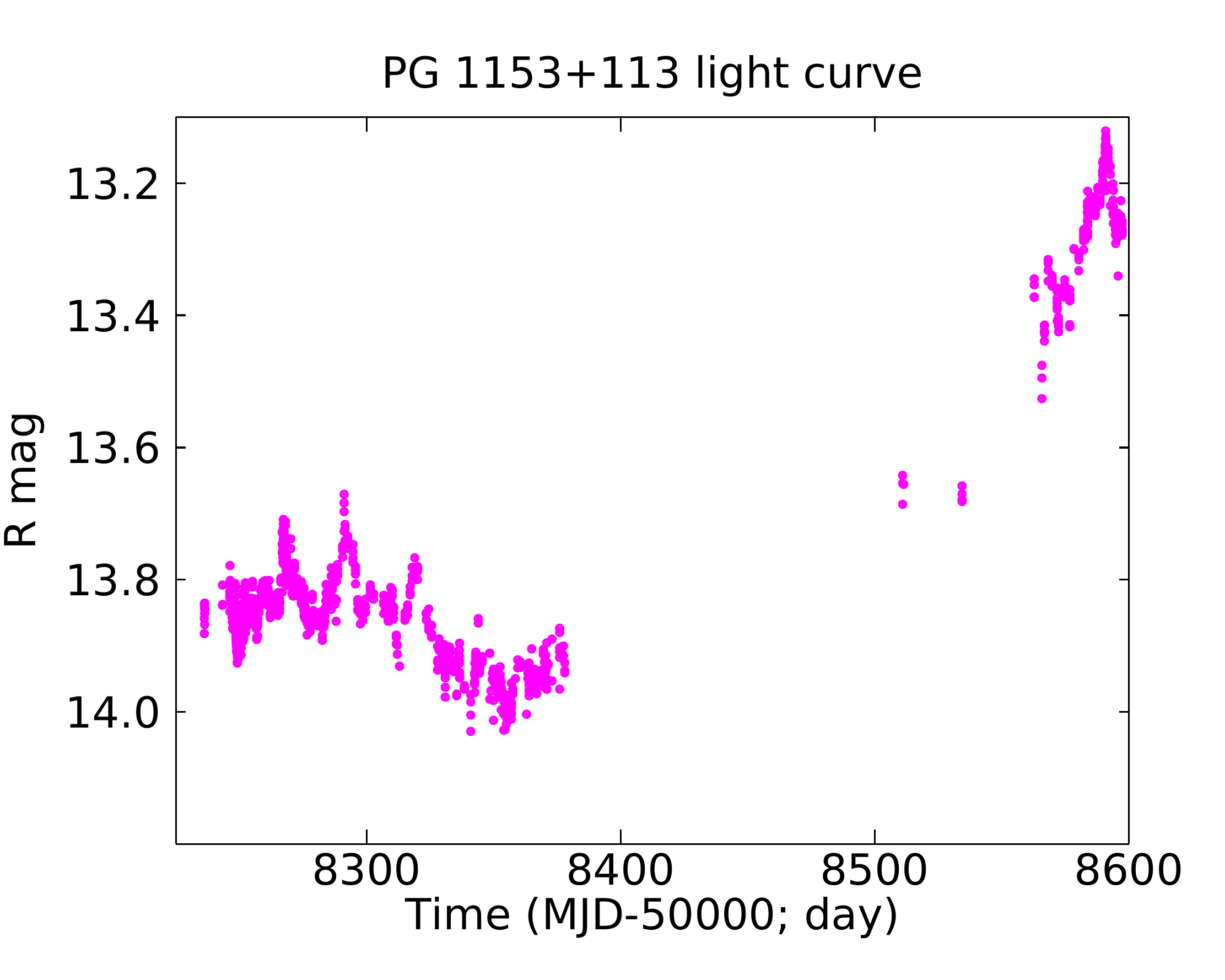}}\\
{\includegraphics[width=0.48\textwidth,angle=0]{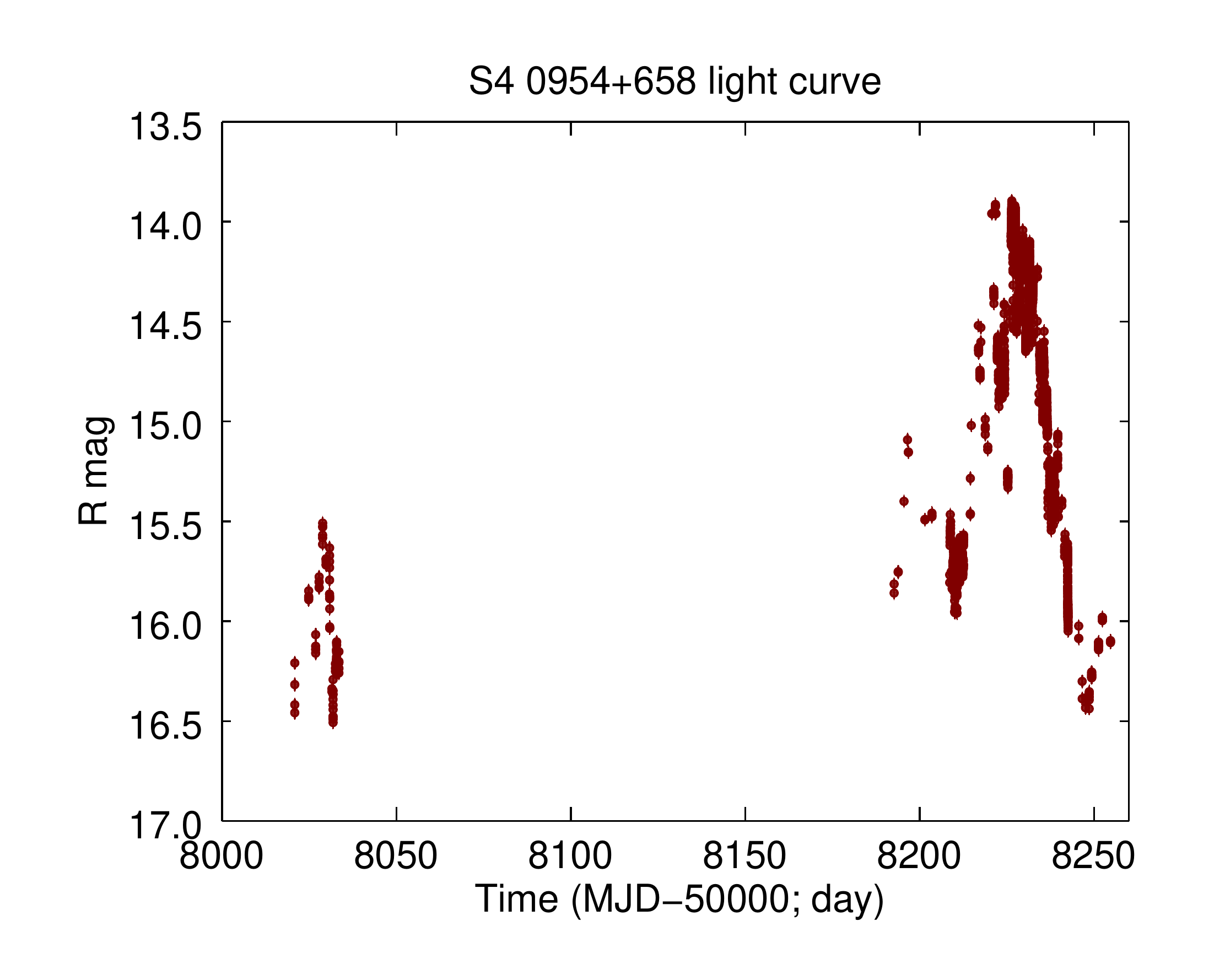}}
\hspace{-0.1cm}
{\includegraphics[width=0.48\textwidth,angle=0]{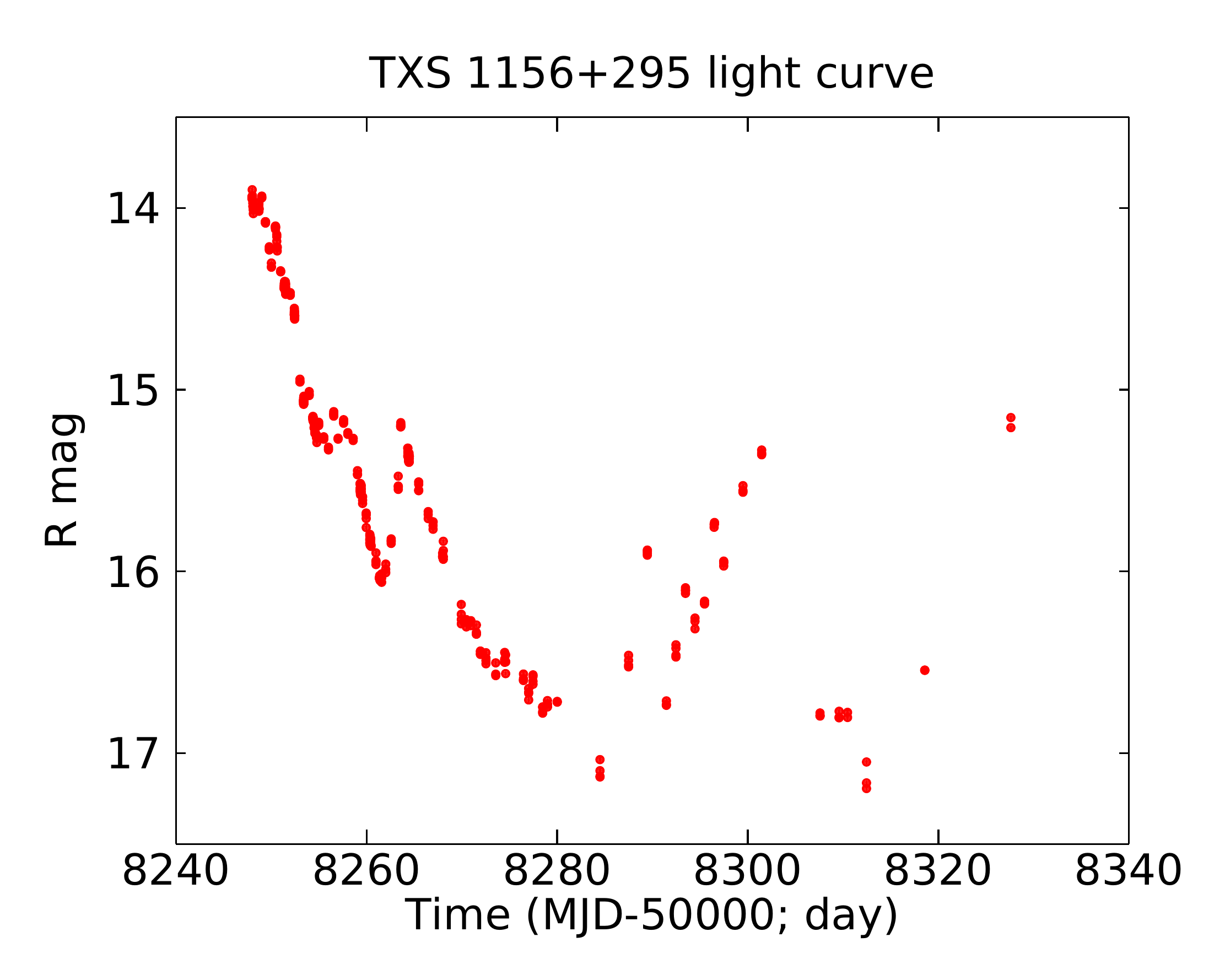}}\\
{\includegraphics[width=0.48\textwidth,angle=0]{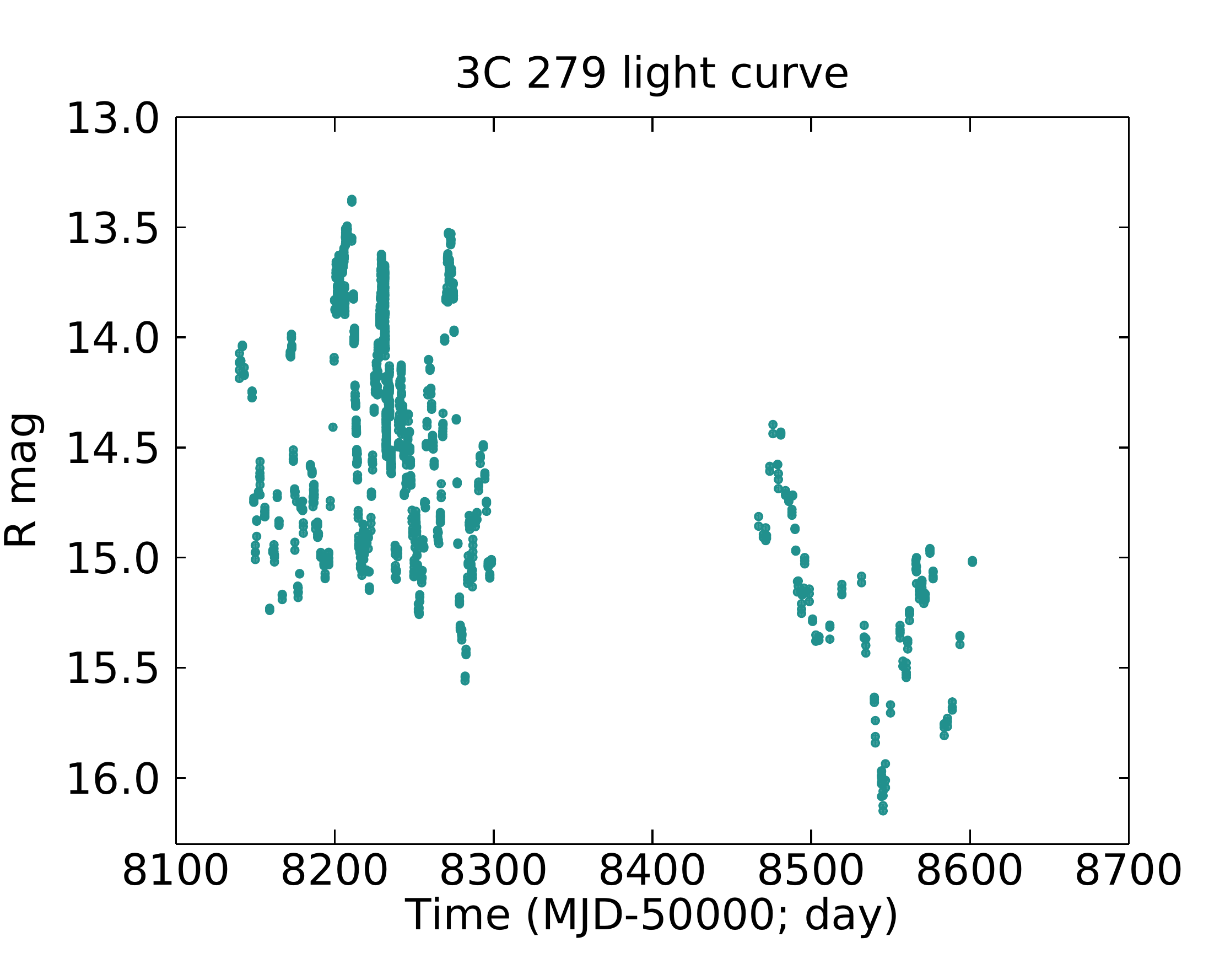}}
\hspace{-0.1cm}
{\includegraphics[width=0.48\textwidth,angle=0]{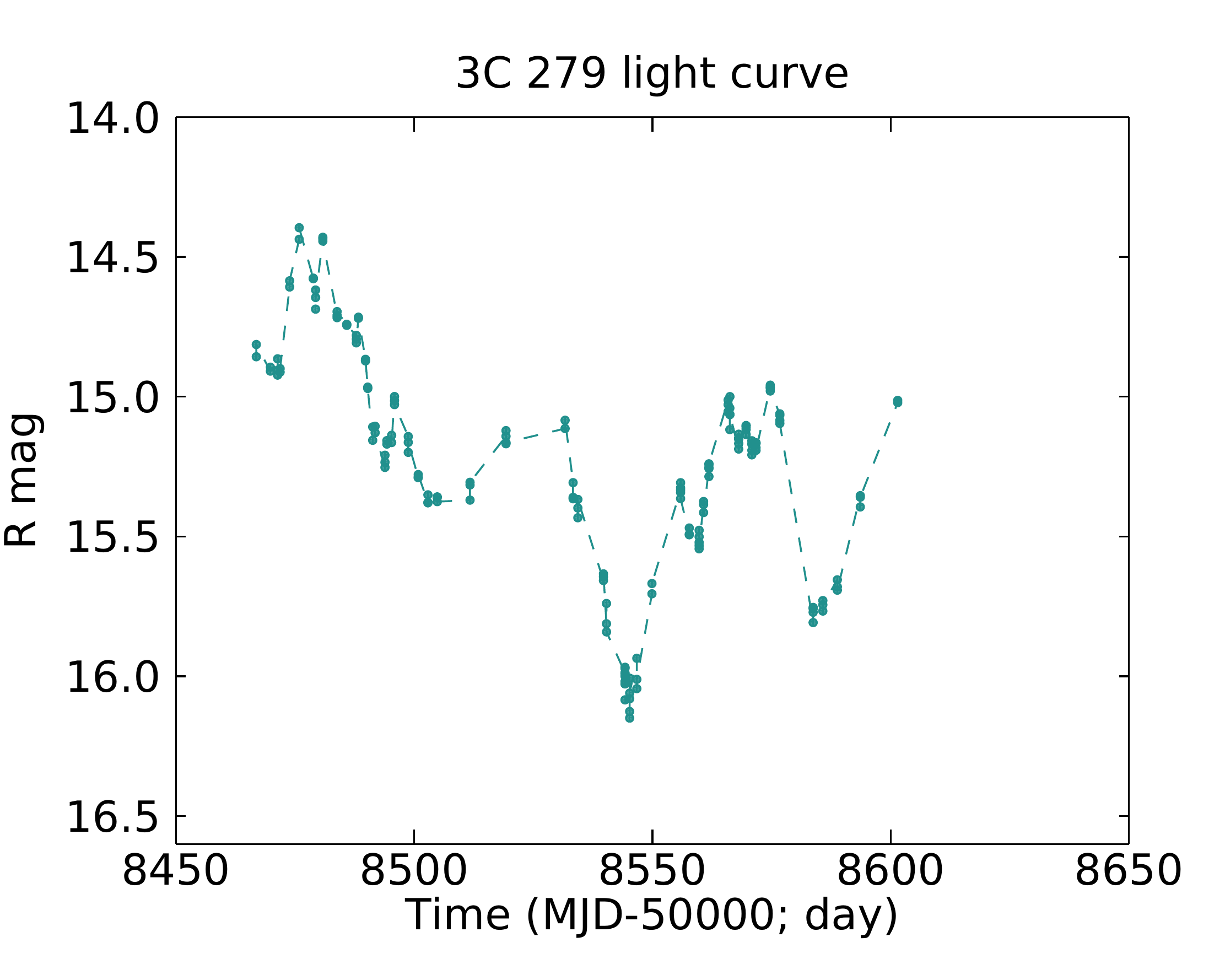}}
\caption{Optical (R-band) observations of the sample of blazars }
\label{fig:1}
\end{center}
\end{figure*}

\section{Source description}
\label{sec:3}
\subsection*{OJ 49}
Blazar OJ 49 is a BL Lacertae object located at a red-shift of 0.17386.
 The Very Large Array (VLA) radio image of the source at 20~cm \citep{Antonucci1985} shows a sharply curved extending jet, while a 43 GHz Very Long Baseline Array (VLBA) image shows a highly polarized jet extending about 0.6 mas from the core \citep{Lister1998}. Similarly, VLBA images at 22 GHz revealed a prominent jet ejecting knotty components at apparently superluminal speeds \citep{Jorstad2001}. During VLBI Space Observatory Program mission, the core size of the 5 GHz radio emission was estimated to be 0.5 mas \citep{Dodson2008}. In the optical and near-IR bands, the blazar exhibits strong intraday variability of polarization and total flux \citep[see][]{Sitko1985,Smith1987}.
 
 \subsection*{S4 0954+658 }
 BL Lac S4 0954+658 is known to display significant flux and polarization variability on both intra-day and longer timescales. Long term optical variability of the source by \cite{1999A&A...352...19R} showed large amplitude-flux modulations. Similarly, \citet{Papadakis2004} made multi-band optical observations of the source for a few nights and found the source to be variable by $<$ 5\% within the observational period. \cite{2014AJ....148...42M} studied the multi-wavelength behavior of S4 0954+658 during a powerful outburst in March-April 2011 using 
 optical (R-band) photometric and polarimetric monitoring and VLBA observations. The authors reported an increase of the flux by 2.8 mag within a period of two months, and a steep intra-night flux increase of $\sim$0.7 mag. Similarly, in a study involving multi-color photometric and polarization observations of the blazar during 2008--2012, the source revealed a power-law spectrum and high degree of polarization, confirming the synchrotron nature of the emission \citep{2015ARep...59..551H}. The source also displayed significant color variability with a trend of bluer-when-brighter (BWB). \citet{2016Galax...4...24M} investigated the behavior of the source during the outburst in early 2015 using observations from cm-wave to $\gamma$-ray energies. It was found that the optical flaring coincided with a similar flare observed in $\gamma$-rays. During the same period, very-high-energy $\gamma$-ray emission was detected and the ejection of a new, bright polarized superluminal knot was observed by the VLBA at 43 GHz. More recently, \cite{Vlasyuk2022} reported the fastest fall in the source's optical flux, by $\sim$ 0.25 mag. within 15 minutes, accompanied by a minute-timescale QPO.

 \subsection*{TXS 1156+295}
Blazar TXS 1156+295, also known as Ton 599 and as 4FGL J1159.5 + 2914 in the Fermi-LAT 4th catalog, is an FSRQ situated at the position of R.A. = $11^h59^m31.8^s$ and Dec. = +29$^{o}14'43''.8$ and lies at a redshift $z = 0.725$ \citep{Hewett2010}. The source was first detected in the $\gamma$-ray band by the Energetic Gamma Ray Experiment Telescope (EGRET), and later was also detected in very high-energy emission ( $>$ 100 GeV) by VERITAS \citep{Mukherjee2017}. In the optical band, the blazar is highly variable in all timescales. \cite{Fan2006} presented a study of the sources using photometric observations which showed a large variation ($\Delta m \sim$ 5.8 mag) in the optical flux on timescales of a few years. In 2017 the source was reported to have undergone optical flaring \citep{Pursimo2017}. In the $\gamma$-ray band, power spectral density and flux distribution analysis of decade-long Fermi/LAT observations was carried out by \citep{Bhatta2020}. More recently, \citet{Rajput2021} studied the flux and spectral variability of the source during its $\gamma$-ray flaring in 2021. Also, variability of emission-line during a non-thermal outburst was reported by \citet{Hallum2022}.

\subsection*{3C 279}
Blazar 3C 279 is a FSRQ source profusely emitting in hard X-ray and $\gamma$-rays. Highly variable across a wide range of spectral bands \citep[see][and the references therein]{Hayashida2015,Paliya2016}, it is one of the few FSRQs detected above 100 GeV \citep{MAGIC2008}. The source reveals a compact, milliarcsecond-scale radio core ejecting radio knots with a bulk Lorentz factor $\Gamma= 15.5\pm 2.5,$ in a direction making an angle $\theta_{obs}=2.1\pm 1.1^{\circ}$ to the line of sight \citep{Jorstad2005,Jorstad2004}. 
During Whole Earth Blazar Telescope (WEBT) campaigns, the source flux in the optical was reported to undergo an exponential-type decay on a timescale of $\sim10$~d \citep{2007ApJ...670..968B}. Similarly, \citet{Larionov2008} in another WEBT campaign in 2006-2007 observed a slower but large flux decline, $\sim 3$ mag on a timescale of $\sim$100~d, in the optical and near-IR band. \citet{Bhatta2018c} discussed the flux and spectral variability properties of the source in the hard X-ray band during intra-day timescales. More recently, \citet{Agarwal2019} presented multi-band optical variability of the source and found that large amplitude variability within a timescale of a few months and a mild BWB trend on shorter time-scales. The optical observations of this source are presented in the bottom panels of Figure \ref{fig:1}. A sharp fall in the flux ($\Delta m \sim 2.0$) within a timescale of 100 days is separately shown in the right panel.

\subsection*{PG 1553+113}
Blazar PG 1553+113 is a BL Lac source which has been studied from radio to $\gamma$-rays in different observation campaigns \citep[e.g.][]{Osterman2006,Ackermann15,Raiteri2015,Raiteri2017}. The source is famous for its $\sim$ 2.18 year periodicity, first revealed in Fermi/LAT observations \citep{Ackermann15}. It was classified as a BL Lac object based on its featureless spectrum (Falomo \& Treves 1990) and was further sub-classified as a high-peaked BL Lac (HBL) object \citep[see][and references therein]{2002A&A...383..410B}. Evidence of very-high-energy $\gamma$-ray emission from this source was first reported by H.E.S.S. in 2005 \citep{2006A&A...448L..19A} and was later confirmed by observations
above 200 GeV with the MAGIC telescope at a significance level of 8.8 $\sigma$ \citep{2007ApJ...654L.119A}. Due to its featureless optical spectrum, the redshift of PG 1553+113 remains highly uncertain. Measurements using the Cosmic Origins Spectrograph onboard the Hubble Space Telescope yielded a lower limit of 0.395 \citep{2010ApJ...720..976D}. The small statistical uncertainties of the VERITAS energy spectrum help constrain the upper limit of the redshift to a value of $\sim$ 0.62 \citep{2015ApJ...799....7A}. During multi-frequency WEBT campaign in 2013 April--August, the source was found to display a general BWB trend in the optical regime \citep{Raiteri2015}. \citet{2018ApJS..237...30M} studied the source using the optical multi-band observations in the yearly timescale. During the observation period the source exhibited moderately varying multi-band emission, without any inter-band lag. \cite{2019ApJ...871..192P} reported variable emission in the optical V and R bands with a mean optical spectral index of $\sim$ 0.83$\pm$ 0.21.

\begin{figure*}
\begin{center} 
{\includegraphics[width=0.32\textwidth,angle=0]{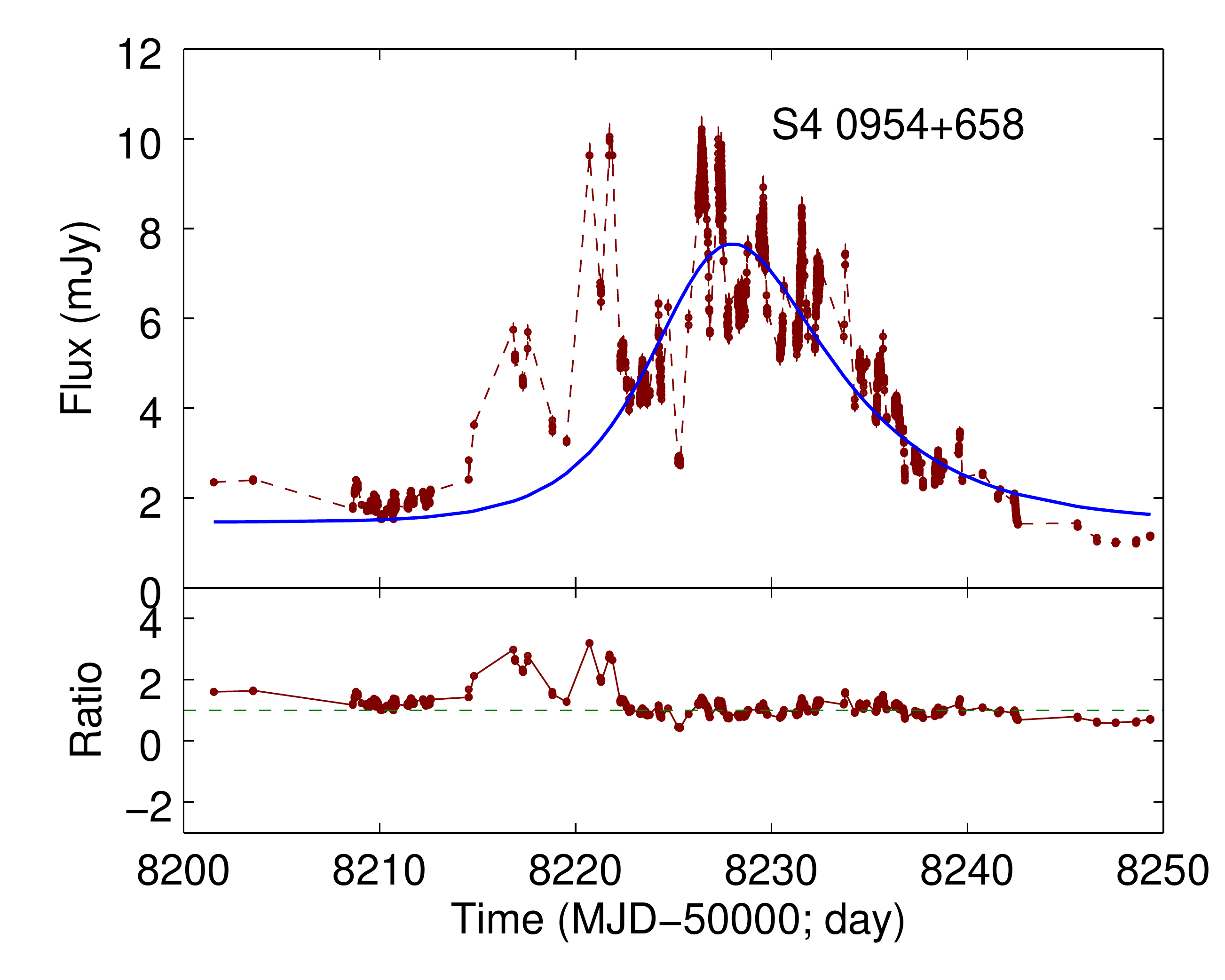}}
{\includegraphics[width=0.32\textwidth,angle=0]{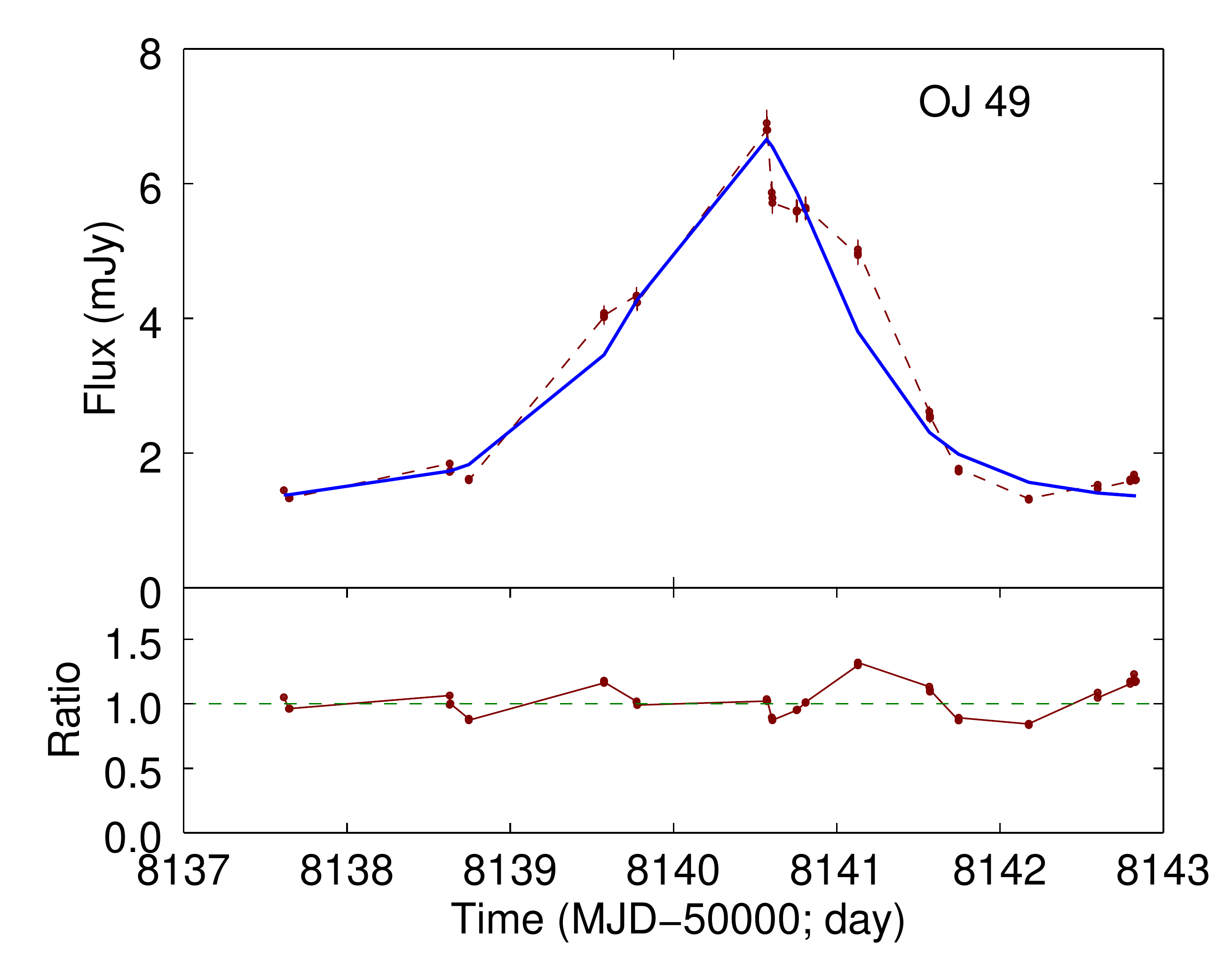}}
{\includegraphics[width=0.32\textwidth,angle=0]{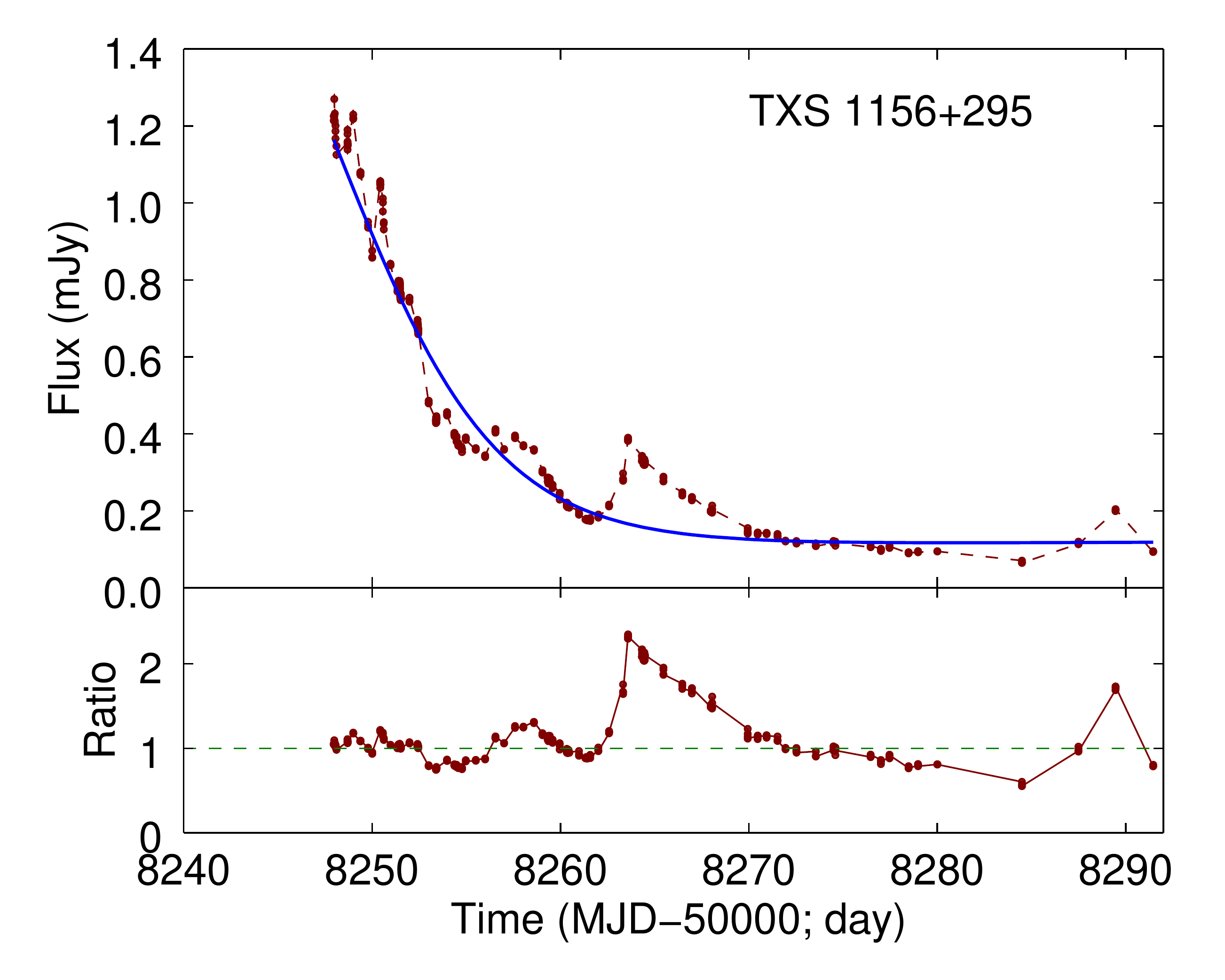}}
\caption{Fitting of the flares observed in optical (R-band) observations of the sample blazars with a curve (shown in blue color) parameterized with an exponential rise and decay are presented in the upper panels of the figures. The lower panels show the ratio between the observations and the models. The corresponding rise and decay times from the best-fitting models are presented in the 8th and 9th columns of Table \ref{table:2}.
 }
\label{fig:2}
\end{center}
\end{figure*}

\section{Analysis and Results}
\label{sec:4}
In order to characterize the properties of the flaring episodes in the target sources, we analyzed the optical observations through multiple analysis methods. The methods and results of the analyses for the individual sources are presented below.

\subsection{Variability measures and timescales}
In order to obtain a quantified measure of the observed variability in the sources, the following three measures of variability are estimated. Variability amplitude (VA) as given in \cite{Heidt1996} provides an measure of the net flux change during the observation period and is written as
\begin{equation}
\rm VA=\sqrt{\left ( A_{max}-A_{min} \right )^2-2\sigma ^2}
\end{equation}
where $A_{max}$ and $A_{min}$ are the maximum and minimum of the source magnitude, respectively, and $\sigma$ represents mean error in the magnitude measurements. The VA of the sample sources are listed in Table \ref{table:2}. Subsequently, the VA in magnitudes then can be directly converted into flux ratio using the relation $f_1/f_2=10^{-0.4(m_1-m_2)}$, where $m_1$ and $m_2$ are the initial and final magnitudes of the variable source. This allows us to compute the factor by which the flux changed during the observational period.

 We observed high-amplitude rapid variability in the blazars included in our study. As seen in Figure \ref{fig:1}, and also indicated by the VA (in magnitudes) listed in the 5th column of Table \ref{table:2}, the optical flux of blazar 3C 279 was observed to change by $\sim13$ times within a timescale of  a few hundred days. Similarly, in OJ 49 we measured a flux change by a factor of $\sim$12 within the period of $\sim12$ days as shown in the top panel of Figure \ref{fig:1}, and in TXS 1156+295 by a factor of $\sim$16 within a timescale of $\sim$22 days as seen in the middle panel of Figure \ref{fig:1}. Likewise, the optical light curve showing flux modulation in the blazar PG 1553+113 is presented in the top right panel of Figure \ref{fig:1}.

The VA considers only extreme values and therefore provides a measure for the peak-to-peak magnitude change. Average variability during the entire period can be quantified by estimating their fractional variability (FV) given as
\begin{equation}
F_{var}=\sqrt{\frac{S^{2}-\left \langle \sigma _{err}^{2} \right \rangle}{\left \langle F \right \rangle^{2}}} ,
\end{equation}
where $S^{2}$ and $\left \langle \sigma _{err}^{2} \right \rangle$ represent the variance and the mean of the squared measurement errors, respectively; and the uncertainty in FV can be expressed as
\begin{equation}
\centering
\sigma_{F_{var}}=\sqrt{ F_{var}^{2}+\sqrt{ \frac{2}{N}\frac{\left \langle \sigma _{err}^{2} \right \rangle^{2}}{\left \langle F \right \rangle^{4}}+ \frac{4}{N}\frac{\left \langle \sigma _{err}^{2} \right \rangle }{\left \langle F \right \rangle^{2}} F _{var}^{2}}} - F_{var}
\end{equation}
 (\citealt{Vaughan2003}, see also \citealt{Bhatta2018a}). The FV values for the sample sources are listed in the 6th column of Table \ref{table:2}, which show large large average flux variability during the period.

 The variability timescale ($\tau_{var}$) can be taken as the e-folding timescale of flux change given by
 \begin{equation}
 \tau_{var}= \left | \frac{\Delta t}{ \Delta lnF} \right |,
 \end{equation}
\citealt[][]{Burbidge1974}, see also \citealt{Bhatta2018c}), where $\Delta t$ is the time interval corresponding to the change in natural logarithm of flux measurements. Using the above relation, $\tau_{var}$ of the sample light curves are calculated and listed in the 7th column of Table \ref{table:2}. Of the sample sources, a shortest $\tau_{var}$ of 11 minutes is observed blazar 3C 279, whereas the source PG 1553+113 appears to show relatively slower variability with a $\tau_{var}$ of 65 minutes.

 Rapid variability in shorter timescales can be associated with the synchrotron cooling timescales, which can be expressed as,
 \begin{equation}
 t_{cool}\sim 7.74\times 10^{8}\gamma^{-1} B^{-2} \ s
 ,\end{equation}
 \noindent where $\gamma$ and B represent electron Lorentz factor and ambient magnetic field, respectively. Also, we use $\beta \sim 1$ considering ultra-relativistic electrons \citep[see also][]{Bhatta2018c}. Using a typical jet magnetic field of 1 Gauss and a minimum variability timescale of 30 minutes, the energy of the relativistic electrons emitting optical synchrotron emission can be estimated to $\sim 3\times10^5$ Lorentz factors. Moreover, following the causality argument, the timescale $\tau_{var}$ can be used to estimate the upper limit for the minimum size of the emitting region ($R$) as given by $R\geq\frac{\delta }{\left ( 1+z \right )}c\tau_{var}$; where $\delta$, Doppler factor, is defined as $\delta =(\Gamma \left ( 1-\beta cos\theta \right ))^{-1}$, and for the velocity $\beta=v/c$ the bulk Lorentz factor can be written as $\Gamma=1/\sqrt{1-\beta^{2}}$. If we use a typical value of Doppler factor of 10 and $z=0.5$, a 30-minute minimum variability timescale could have arisen from a region that is 1/1000-th of a parsec in scale.

\begin{table*}
 \caption{Optical observations of a sample of blazars and their variability properties}
 \centering
 \label{table:2}
 \begin{tabular}{l|r|l|l|c|c|c|c|c}
 \hline
 Source name & Duration (d) &Npt.& mean mag& VA (mag)&Fvar (\%)&t$_{\rm var}$ (min.)&t$_{\rm r}$ (d)&t$_{\rm d}$ (d)\\
 (1)& (2) &(3)& (4)& (5)&(6)&(7)&(8)&(9)\\
 \hline
 OJ 49 &164.28& 620& 16.28 & 2.75&66.23$\pm$0.24&38.24$\pm$11.80&0.57&0.44\\
 TXS 1156+295 &79.65 & 650 & 17.45 & 3.30&62.78$\pm$0.13 &19.06$\pm$13.73&-&4.10\\
 PG 1553+113 &97.43 & 1071& 15.57& 0.91 & 23.61$\pm$0.21&64.76$\pm$11.18&-&-\\
 3C 279 &461.58&1831 & 17.59 & 2.78&46.16$\pm$0.50 &11.73$\pm$7.80&-&-\\
 S4 0954+658 &242.70& 2988 & 14.80 & 2.61 &47.60$\pm$0.17&17.10$\pm$6.18&3.05&5.15\\ 
 \hline
 \end{tabular}
\end{table*}

\subsection{Symmetry Analysis: Rise and Decay Profiles}
The rise and decay profiles of flares in blazar light curves can be associated with the particle acceleration and cooling timescales, respectively, and thereby can be linked to the physical processes leading to the flaring episodes. To characterize the flaring properties, we performed symmetry analysis of the flares. For that purpose, we took parts of the light curves of the sources OJ 49, S4 0954+658, and TXS 1156+295 (see Figure \ref{fig:1}) which showed well resolved flares, with a well-defined generally monotonic rise and decay, and the three flares were fitted by a functional form representing the temporal structure of the exponential rise and decay described by
\begin{equation}
\label{symm}
F\left ( t \right )=F_{c}+F_{0}\left [ e^{\frac{t_{0}-t}{t_r}}+ e^{\frac{t-t_{0}}{t_d}}\right ]^{-1},
\end{equation}
where $F_{c}$ is the constant flux level, $F_{0}$ is the amplitude of the flaring structure and $t_{0}$ the center of the flare, and $t_r$ and $t_d$ are the rise and decay times of the flares \citep{2010ApJ...722..520A}. The parts of the light curves of the sources which show distinct flares and the corresponding functional fits are shown in Figure \ref{fig:2}. The rise and decay timescales determined from the fitting are listed in the 8th and 9th columns, respectively, of Table \ref{table:2}. Using the obtained timescales, a symmetry parameter written as $\xi =\left ( t_d -t_r \right )/\left ( t_d +t_r \right )$ can be defined within [-1,1] such that $\xi =-1, +1$ represent completely right- and left-asymmetric flares, whereas $\xi =0$ represents completely symmetric flares. The asymmetry parameters for OJ 49 and S4 0954+658 $\xi =0.13$ and $\xi =-0.25$, respectively. The result indicates that the flares in these sources are, respectively, slightly left- and right-asymmetric. Since the source TXS 1156+295 was only observed during its decay phase its $\xi$ could not be estimated. Also, although the amplitude of the flares in OJ 49 and S4 0954+658 are similar, with normalized amplitudes ($F_{0}/F_{c}$)=8.5 and 8.2 respectively, the rise and decay times are considerably shorter in OJ 49, indicating fast flaring events. In the case of TXS 1156+295 an even faster decay of flux, by 11.3 normalized amplitudes, is observed within 4.3 days.


\subsection{ Periodicity Analysis }
 QPOs in blazars with characteristic timescales of a few years have been frequently reported (\citealt{bhatta16c}, see also \citealt{Zola2016}). However, not many blazar QPOs are observed on timescales of a few days or weeks. We searched for the possible periodic flux modulations in the optical bands using Lomb-Scargle method \citep{Lomb76,Scargle82}. The method modifies the conventional discrete Fourier periodogram such that the least-square fitting of sine waves of the form $X_{f}(t)= A \cos\omega t +B \sin\omega t$ to the data is minimized. The periodogram is given as
\begin{equation}
P=\frac{1}{2} \left\{ \frac{\left[ \sum_{i}x_{i} \cos\omega \left( t_{i}-\tau \right) \right]^{2}}{\sum_{i} \cos^{2}\omega \left (t_{i}-\tau \right) } + \frac{\left[ \sum_{i}x_{i} \sin\omega \left( t_{i}-\tau \right) \right]^{2}}{\sum_{i} \sin^{2}\omega \left( t_{i}-\tau \right)} \right\} \, ,
\label{modified}
\end{equation}
where $\tau$ is given by $\tan\left( 2\omega \tau \right )=\sum_{i} \sin2\omega t_{i}/\sum_{i} \cos2\omega t_{i} $\,.

\begin{figure*}
\includegraphics[angle=0,width=0.48\textwidth]{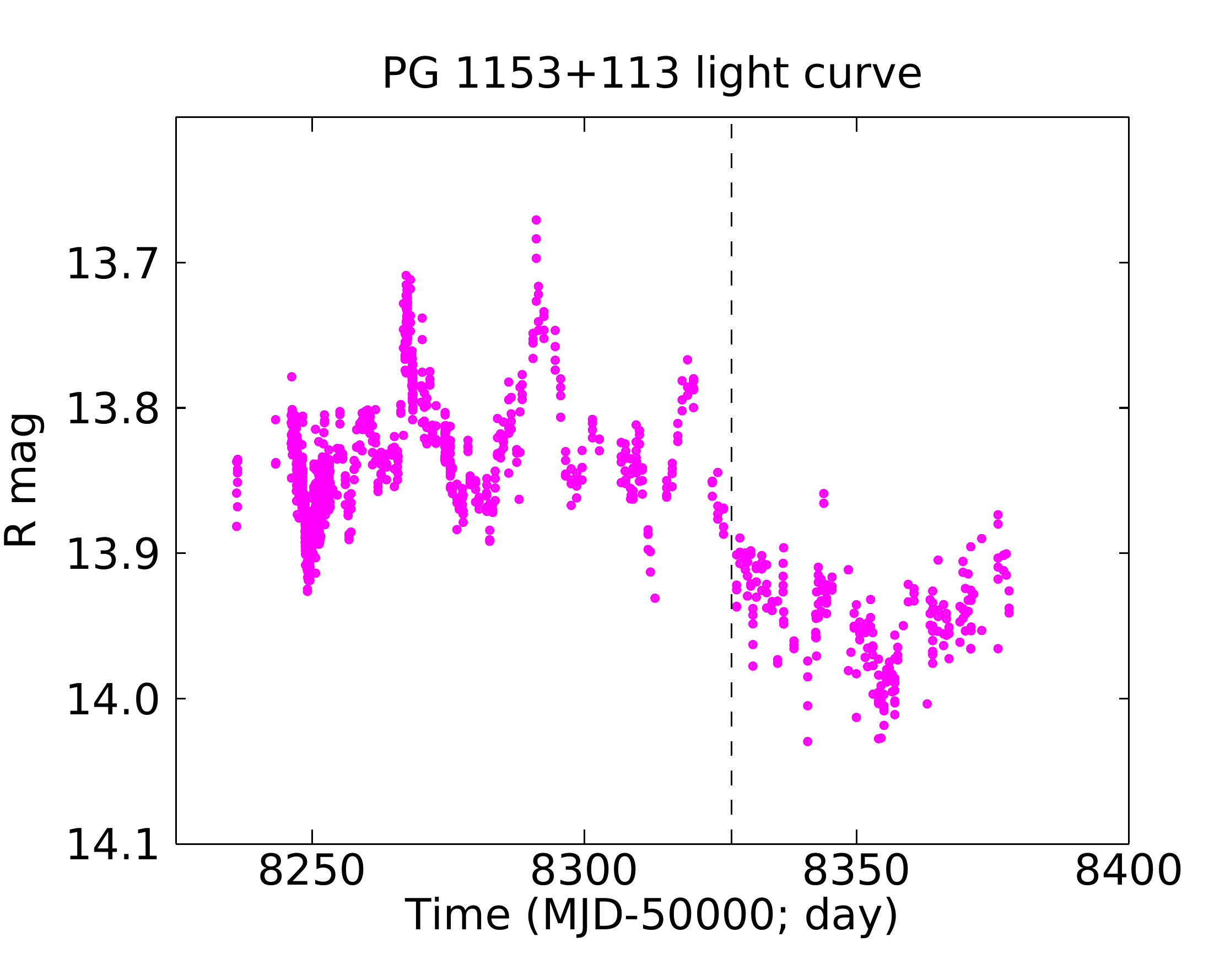}
\includegraphics[angle=0,width=0.45\textwidth]{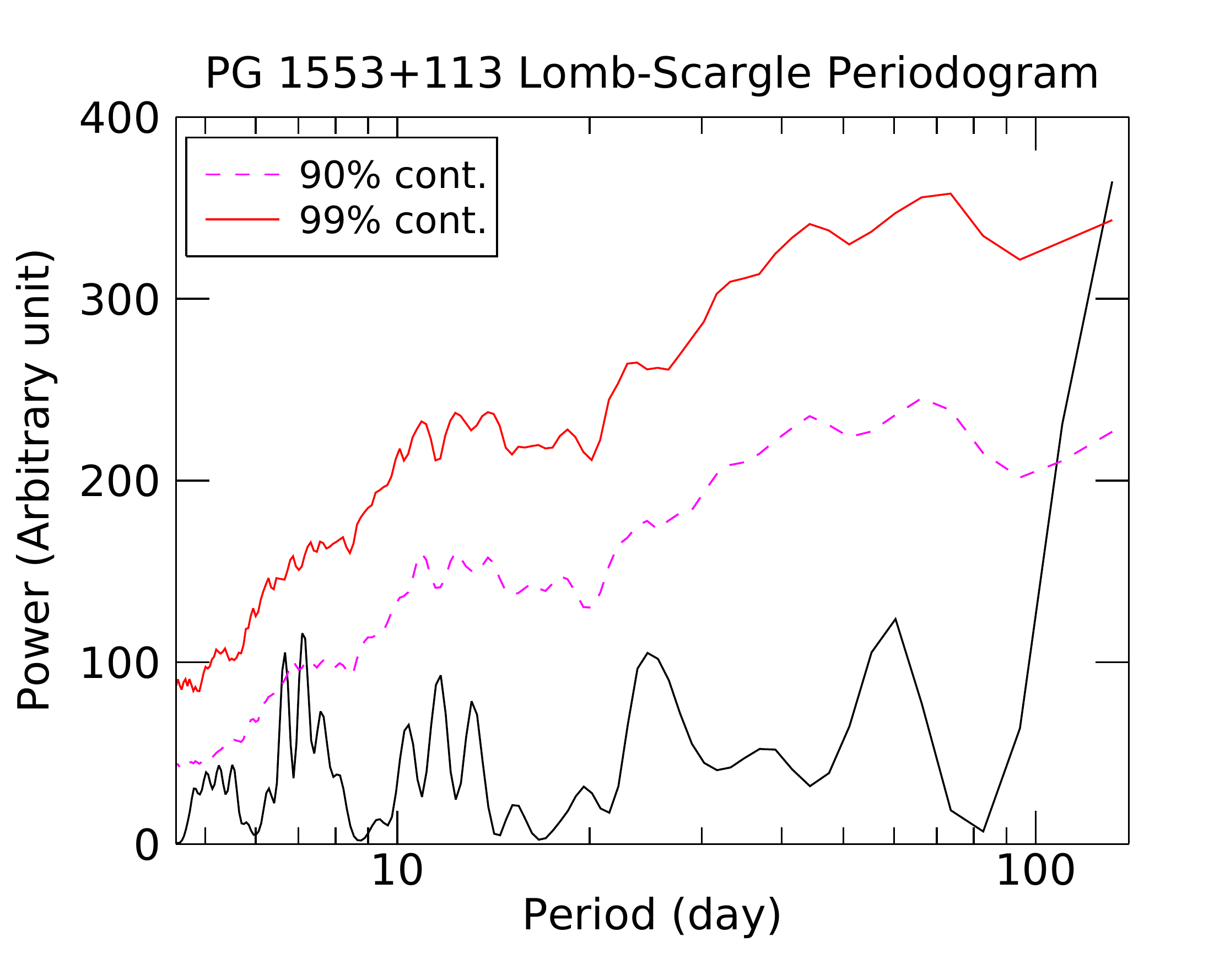}\\
\includegraphics[angle=0,width=0.47\textwidth]{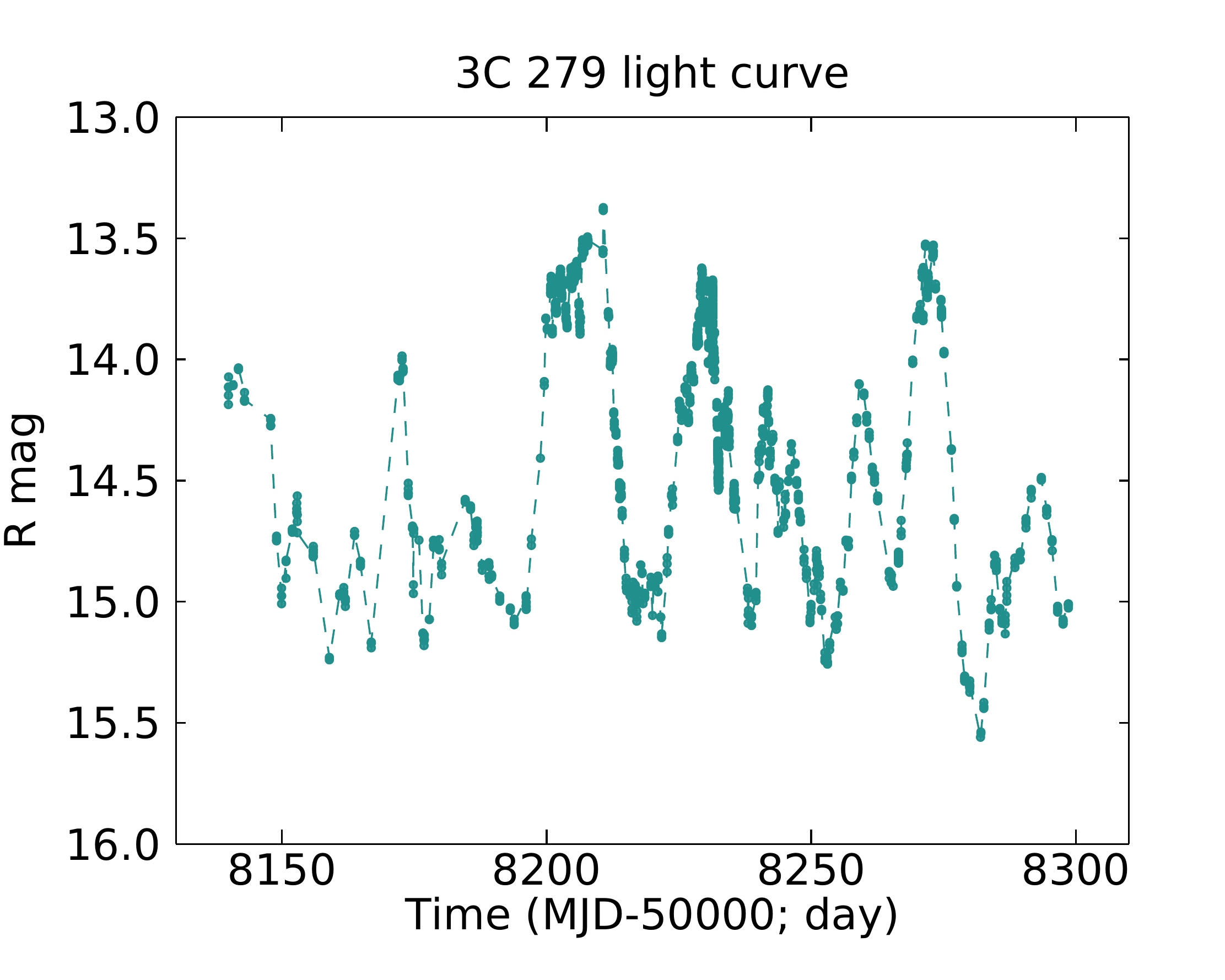}
\includegraphics[angle=0,width=0.47\textwidth]{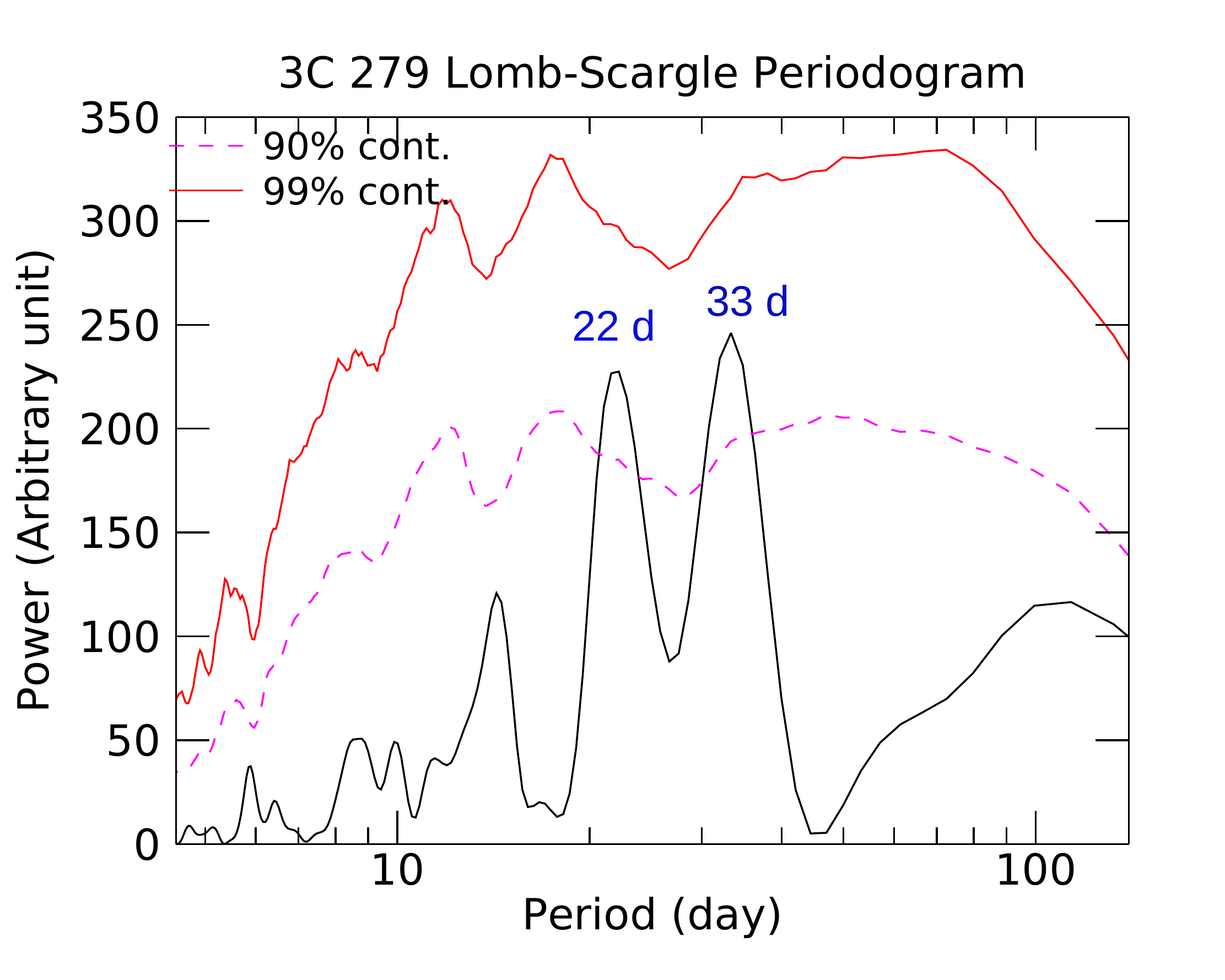}
 \caption{ Lomb-Scargle periodogram and the R band light curve of the blazars PG 1553+113 and 3C 279 are shown in the top and bottom panels, respectively. The 90 and 99\% significance contours from simulation are shown on the LSP diagram by magenta and red curves, respectively.}
\label{Fig4}
\end{figure*}

\begin{figure}
\begin{center}
\includegraphics[angle=0,width=0.50\textwidth]{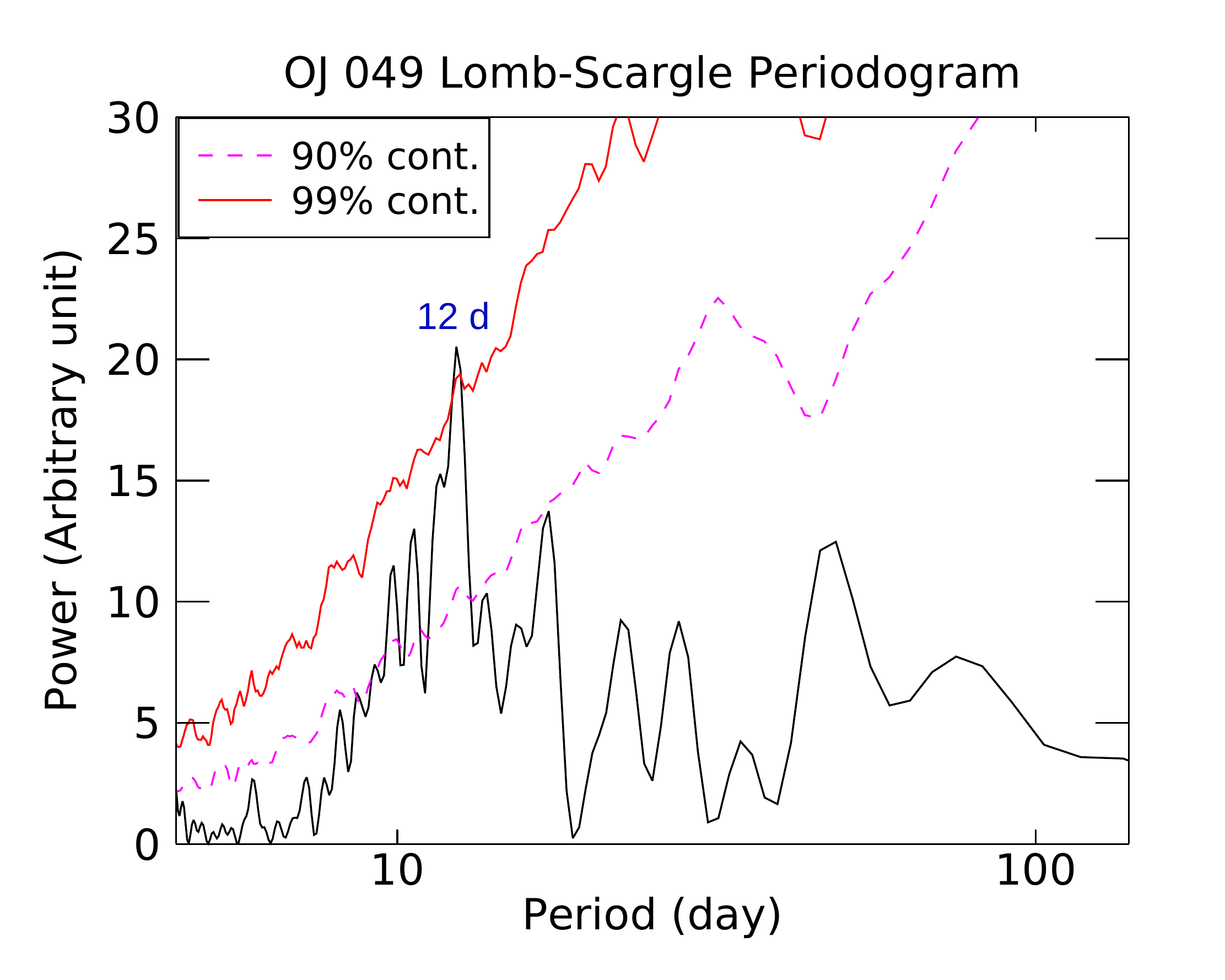}
 \caption{ Lomb-Scargle periodogram of the blazar OJ 49 showing possible $12\pm1.5$ d periodicity. The 90 and 99\% significance contours from simulation are shown by the magenta and red curves, respectively}
\label{Fig5}
\end{center}
\end{figure}

The Lomb-Scargle periodogram (LSP) of the source light curves of PG 1553+113 and 3C 279 are presented in the right panels of Figure \ref{Fig4} and similarly LSP of the source OJ 049 is presented in Figure \ref{Fig5}. The light curve of blazar PG 1553+113 shows peaks at the timescales of $7\pm0.8$ ,$25\pm3$, and $59\pm7$ d, which possibly may be the signs of quasi-periodic oscillations. Similarly the LSP of the blazar 3C 279 shows two prominent peaks around the timescales of 22 and 33 days. Furthermore, in the LSP of the source OJ 049 a prominent peak around 12 day time scale is observed. 

To compute the significance of the observed periodogram features in the source LSP, a large number of light curves were generated by Monte Carlo simulations and their statistical properties were utilized. In particular, the source periodograms based on Discrete Fourier Transform (DFT) were linearly fit in the logarithmic frequency space to obtain a best representative model power spectral density (PSD) using the method described in \citet{Vaughan2005}. To avoid artifacts associated with DFT, some of the data were linearly interpolated to create a evenly spaced sampling. Using the best-fit PSD model, 10~000 light curves mimicking the observations in duration and sampling rate were simulated and the distribution of these simulated LSP periodograms was used to determine 90 and 99\% significance contours \citep[for detail see][]{bhatta16c, Bhatta2020}, shown in magenta and red curves in the corresponding LSP figures. It can be seen that only the 12 d period in the source OJ 049 appears at over the 99\% significance level, although the 7 d period in PG 1552+113, and 22 and 33 d periods in the source 3C 279 are clearly observed. We note that in another blazar Mrk 501 a QPO of a similar timescale of 23 d has been reported previously \citep[see][]{Rieger2000}.

\section{ Discussion}
\label{sec:5}
Blazars are found to be violently variable over all timescales from a few minutes to decades. Apart from stochastic variability often represented by power-law PSD, blazar light curves frequently show MWL flaring events characterized by well-resolved trends of monotonic rise or decline of flux which can last from a few weeks to months. Studies by several blazar monitoring groups are particularly focused on flares in gamma-rays, which are frequently associated with the ejection of radio knots visible in VLBA images \citep{Agudo2011,Wehrle2012,Jorstad2013,Morozova2014}. There is growing evidence that flaring events observed in the gamma-ray wavelength range could be associated with superluminally-moving features crossing stationary features along the jet; it has been found that a large number of flares occur in coincidence with the passage of superluminal knots through the millimeter-wave core\citep[see][and the references therein]{Jorstad2017}. 

Flares can be explained in terms of the sudden enhancement of the blazar flux followed by a strong energy dissipation event. The particles are accelerated to high energies via a number of particle acceleration mechanisms. In the shock diffusive acceleration model, shock waves compress and order the upstream magnetic field, causing the particles to accelerate through the Fermi acceleration mechanism and subsequently leading to the formation of outbursts of synchrotron emission \citep{Blandford1987,Hughes98}. Similarly, in the turbulent jet models the main jet can be thought of being divided into a large number of sub-volumes moving relativistically in random directions \citep[see e.g.][]{Marscher14,Narayan2012}. In this scenario, the passage of shock waves in the turbulent flow of relativistic plasma can heat the particles by compressing the plasma, and subsequently depending upon the size and direction of the motion one single turbulent cell can beam dominantly to appear as a flare. In the case that the jet is highly magnetized, the turbulence can trigger intermittent magnetic instabilities, such as kink instabilities \citep{Spruit2001} or inversions of magnetic field near the base of the jet \citep{Giannios2019}, which lead to magnetic reconnection events and consequently acceleration of the particles to high energies. \citep{Guo2014a,Guo2014ApJ...794..153G}. Similarly, magnetic reconnection events lead to the formation of plasmoids, which can grow to produce rapid flares with distinct and resolved envelopes. In this study, the flare are found to be slightly asymmetrical, which might suggest that the observed skewness in the flare profiles can be linked to both disturbance passing through the emission region and/or geometric effects such as those resulting from light crossing time. The origin of these flares could be extrinsic and intrinsic in nature, as discussed qualitatively in the following sections.

\subsection{ Source intrinsic scenario}
The flares in blazars can be explained in the context of internal shocks propagating along the blazar relativistic jets. 
 As an illustrative example of blazar flares which last about a week, here we attempt to produce a flare using the method worked out by \citet[][]{Kirk1998}. This provides a framework for time-dependent analysis of the homogeneous single zone leptonic model, in which variable emission leading to large flares in the source flux is expected owing to gradual particle acceleration at the shock front and subsequent radiative cooling at the emission region. The model assumes that the shocks are propagating along a cylindrical jet aligned near the line of sight, and that particles injected at a constant rate at the shock wavefront lose energy primarily via synchrotron emission.

 \begin{figure*}
\centering
\includegraphics[angle=0,width=0.40\textwidth]{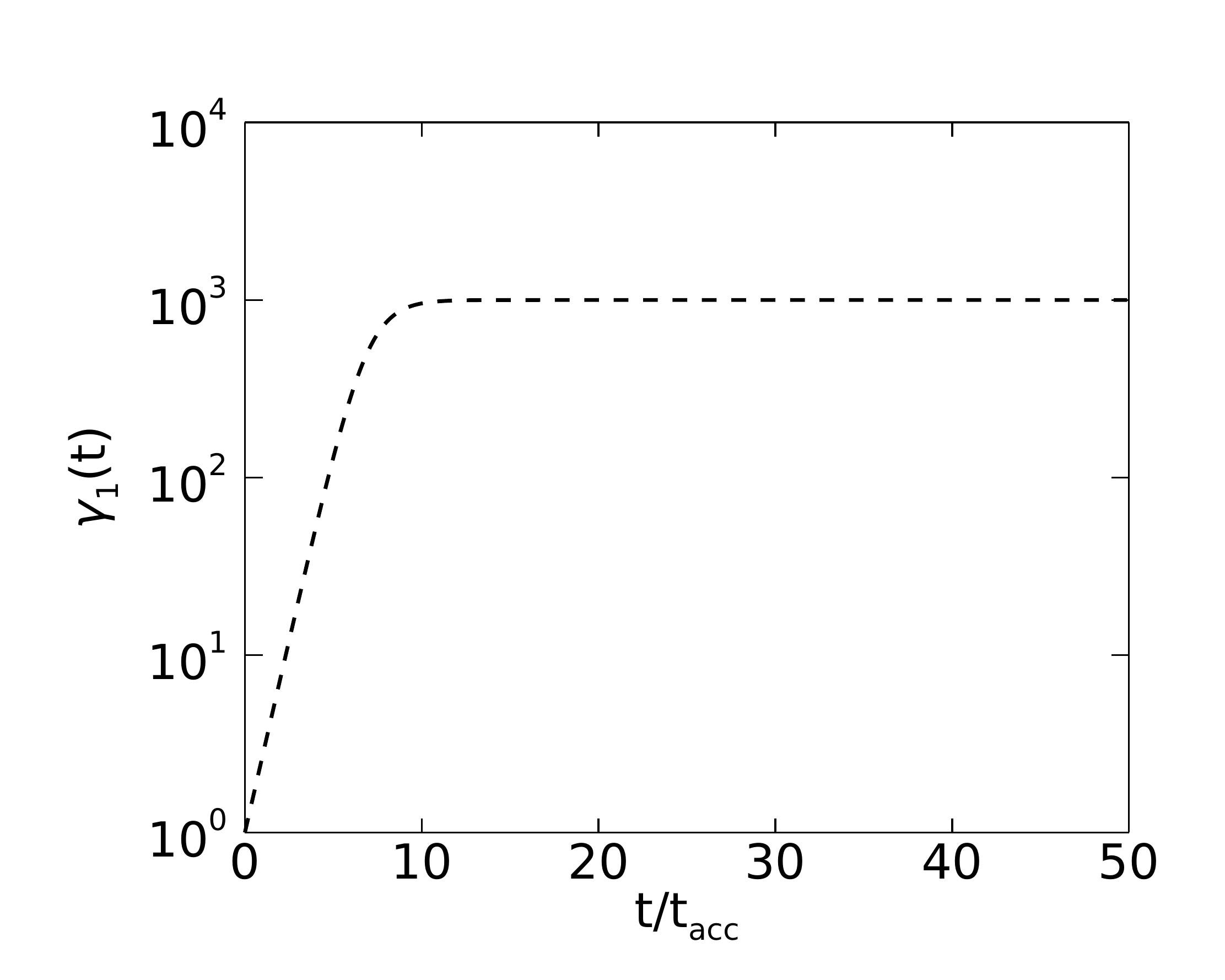}\includegraphics[angle=0,width=0.40\textwidth]{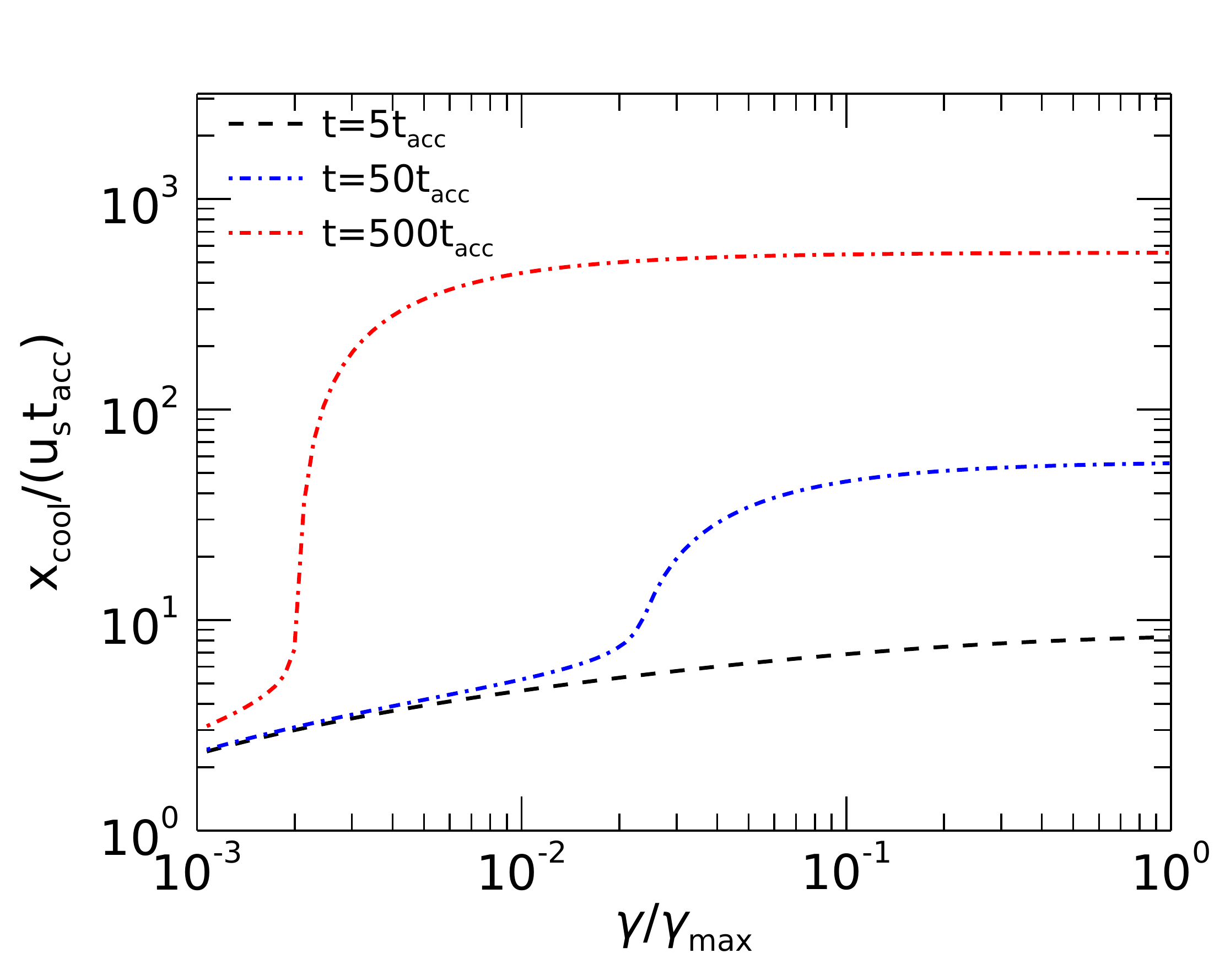}
\includegraphics[angle=0,width=0.40\textwidth]{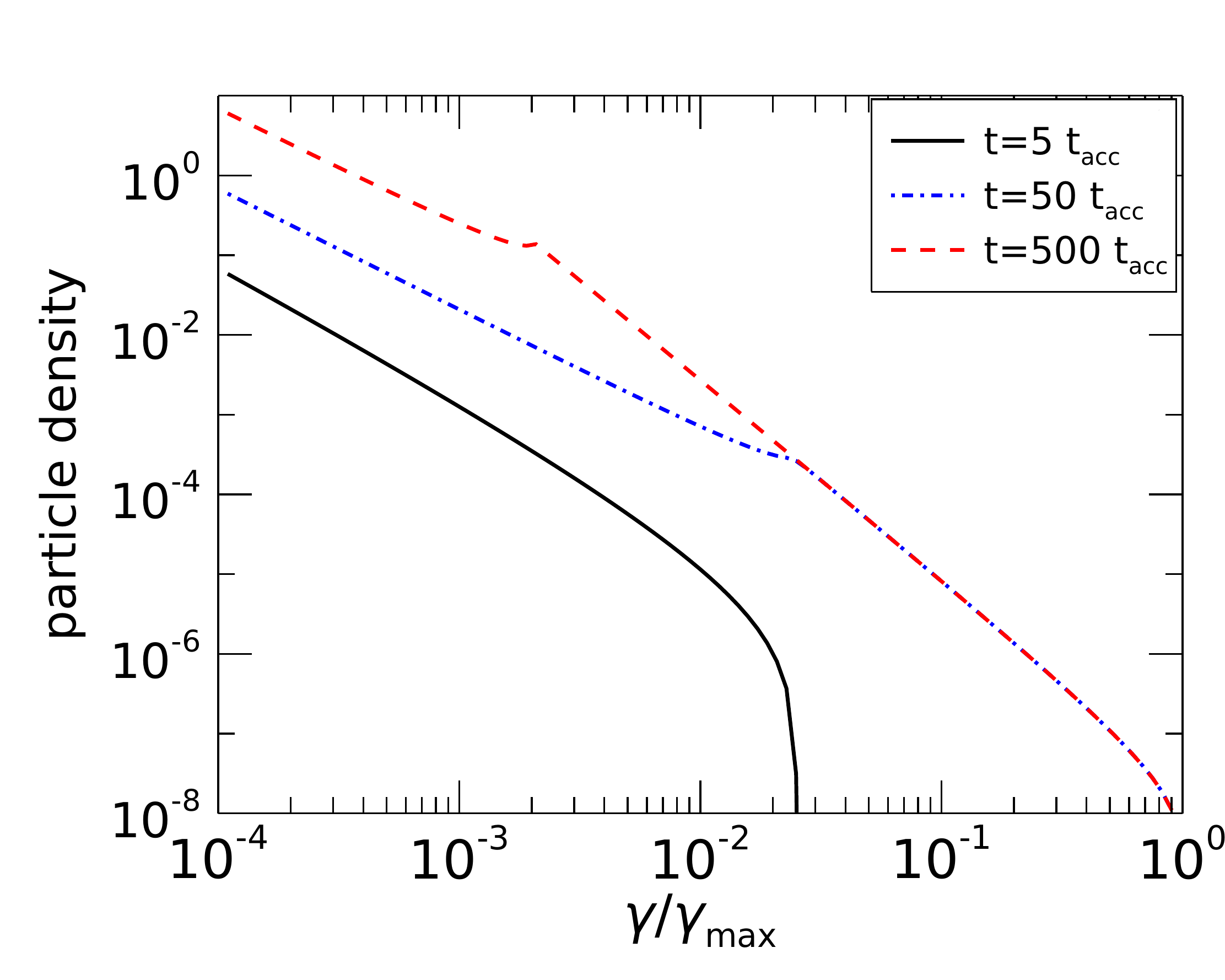}\includegraphics[angle=0,width=0.40\textwidth]{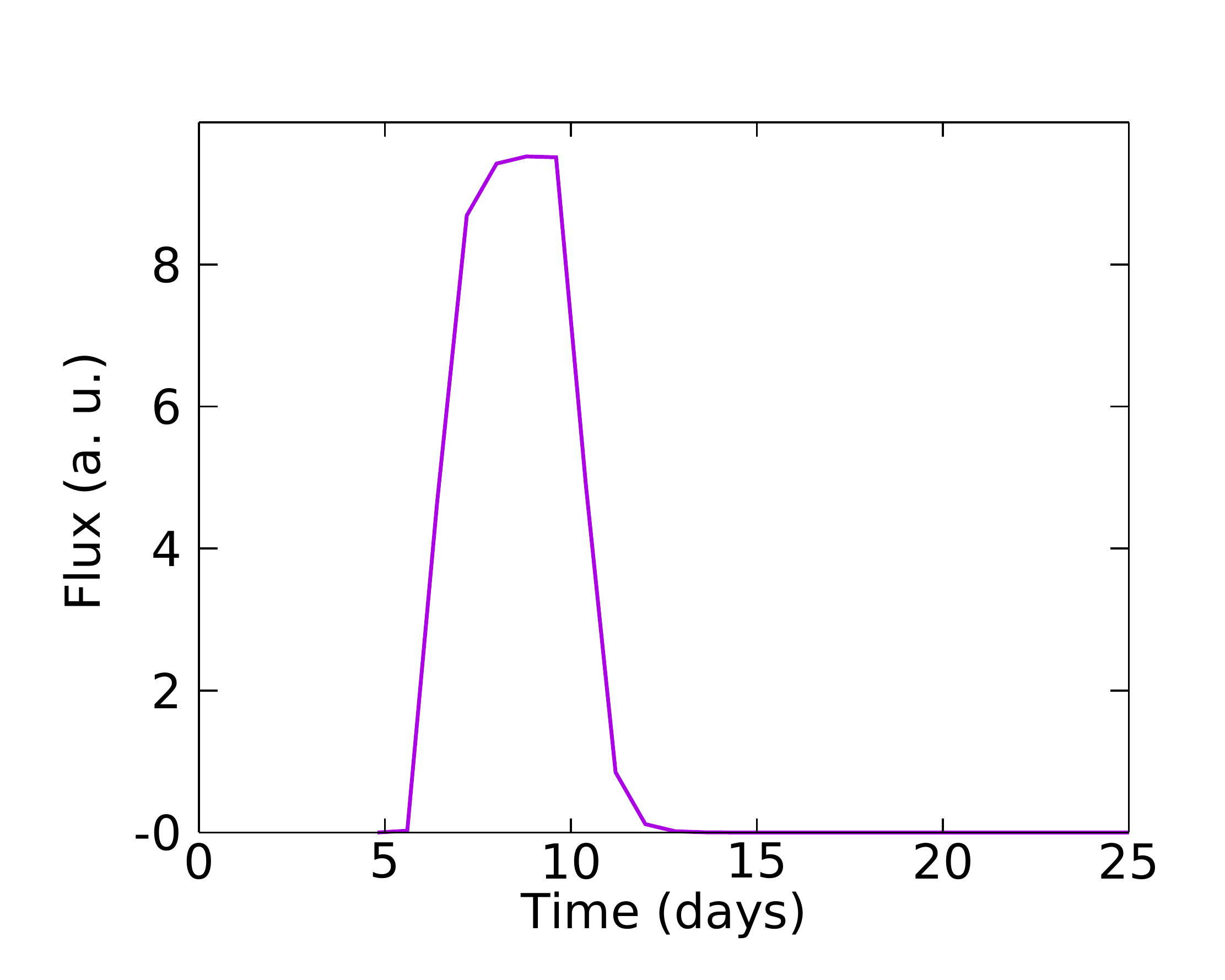}
 \caption{{\it Top}: Time evolution of maximum electron Lorentz factor as it accelerates (left panel). Cooling length behind the shock wave front as a function of electron energy for three different times. {\it Bottom}: The normalized integrated particle (electron) density in the acceleration zone for three different times 5 t$_{acc}$, 50 t$_{acc}$ and 500 t$_{acc}$ as shown by the black, blue and red curves, respectively (left panel). Rise and decay of the blazar emission due to particles accelerated at the relativistic shocks front and thereby cooling, respectively (right panel).}
\label{fig:9}
\end{figure*}

The evolution of the particles in the acceleration zone is given by the diffusion equation,
\begin{equation}
\frac{\partial N}{\partial t}+\frac{\partial }{\partial \gamma }\left [ \left ( \frac{\gamma }{t_{acc}}-\beta _{s} \gamma ^{2} \right ) \right ]N +\frac{N}{t_{acc}}=Q\delta (\gamma -\gamma _{0}).
\label{diffusion1}
\end{equation}
where $\beta _{s} \gamma ^{2}$ represents the loss of the energy by the synchrotron radiation with
\begin{equation}
\beta _{s}=\frac{4}{3}\frac{\sigma_{ T}}{m_{e}c^{2}}\left ( \frac{B^{2}}{2\mu _{0}} \right ),
\label{energy}
\end{equation}

\noindent where $\sigma_{ T}$, B and $\mu _{0}$ are the Thompson-scattering cross-section, the magnetic field and the permeability of free space, respectively. 

\noindent Particle acceleration reaching up to $\gamma _{max}=(\beta_{s} t_{acc})^{-1}$ is described by the equation
\begin{equation}
\gamma 1\left ( t \right )=\left ( \frac{1}{\gamma _{max}}+\left [ \frac{1}{\gamma _{0}}-\frac{1}{\gamma _{max}} \right ]e^{-t/t_{acc}} \right )^{-1},
\label{gamaone}
\end{equation}
which shown in the top panel of Figure \ref{fig:9}.
The distributions of the (normalized) particle densities over the particle energies for the three different acceleration times, i. e., 5, 50 and 500 t$_{\rm acc}$ are presented in the bottom left panel of Figure \ref{fig:9} \citep[see similar Figure 1 in][]{Kirk1998}. The breaks which naturally appear in the curves divide the population into particles with energies greater than those which cool within the source and low energy particles which do not cool within the emission region.
 
\noindent The particle enhancement in the regions is described by the equations,
$ Q\left ( t \right )=Q_{0}$ for t$<$0 and t$>$$t_{f}$ and $Q\left ( t \right )=\left ( 1+\eta _{f} \right )Q_{0}$ for 0$<$t$<$ $t_{f}$.

\noindent We have then,
\begin{equation}
\label{flare}
I\left ( \nu ,t \right )=I_{1}\left ( \nu ,t \right )+\eta _{f}\left [ I_{1}\left ( \nu ,t \right )-I_{1}\left ( \nu ,\left ( 1-u_{s}/c \right )t_{f} \right ) \right ].
\end{equation}
For small angles $cos\theta \rm \sim1$, and $\delta \sim 2\Gamma$ such that the relations $I(\nu,t)=\delta^3I(\nu',t') \sim 8\Gamma^3I(\nu',t')$, and $\nu =\Gamma \left ( 1+\beta \right )\approx 2\Gamma \nu' $ are used to transform the source rest-frame quantities (primed) to the observer's frame (unprimed).
We take $\nu=4.55\times10^{14}$ Hz, the frequency corresponding to the mean effective wavelength of R-band (658 nm) and the shocks traveling down the blazar jet with relativistic speeds ($\beta_{s}=0.1$). The resulting normalized intensity profile mimicking the flaring behavior observed in the blazars is presented in the bottom right panel of Figure \ref{fig:9}.

\subsection{ Source extrinsic scenario}

A large flare in the blazar flux can also appear owing to the Doppler boosted emission when the emission regions travel along a curve trajectory in the jets. In such cases, the rest frame flux (${F}'_{{\nu'}}$) is related to the observed flux ($F_{\nu}$) through the equations

 \begin{equation}
\frac{F_{\nu}(\nu)}{{F}'_{{\nu}'} (\nu)}=\delta^{3+\alpha} \quad\text{and}\quad \delta(t)=\frac{1}{\Gamma \left ( 1-\beta cos\theta \right )}
\label{flux}
\end{equation}

 \begin{figure*}
\centering
\includegraphics[width=0.47\textwidth,angle=0]{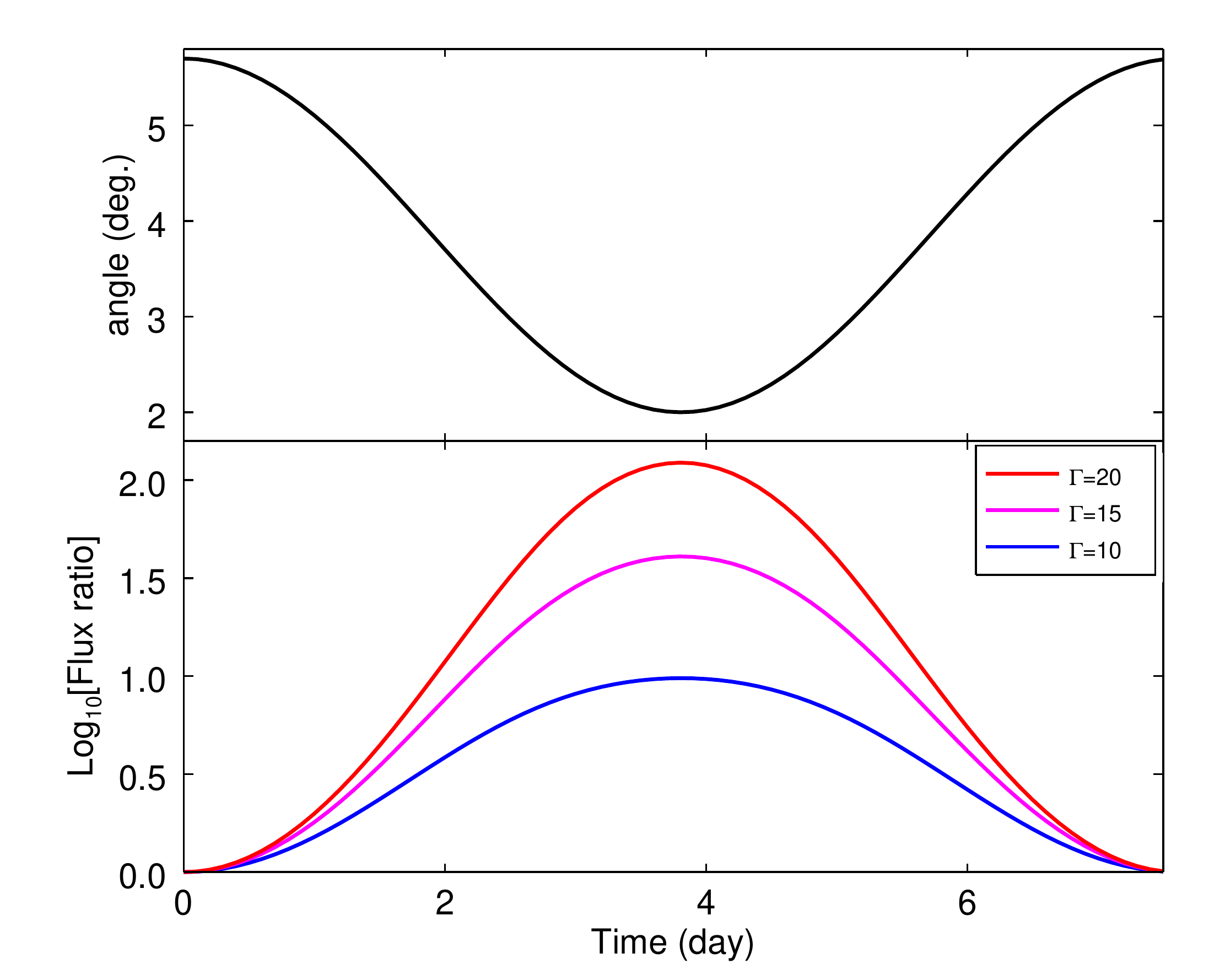}
\includegraphics[width=0.47\textwidth,angle=0]{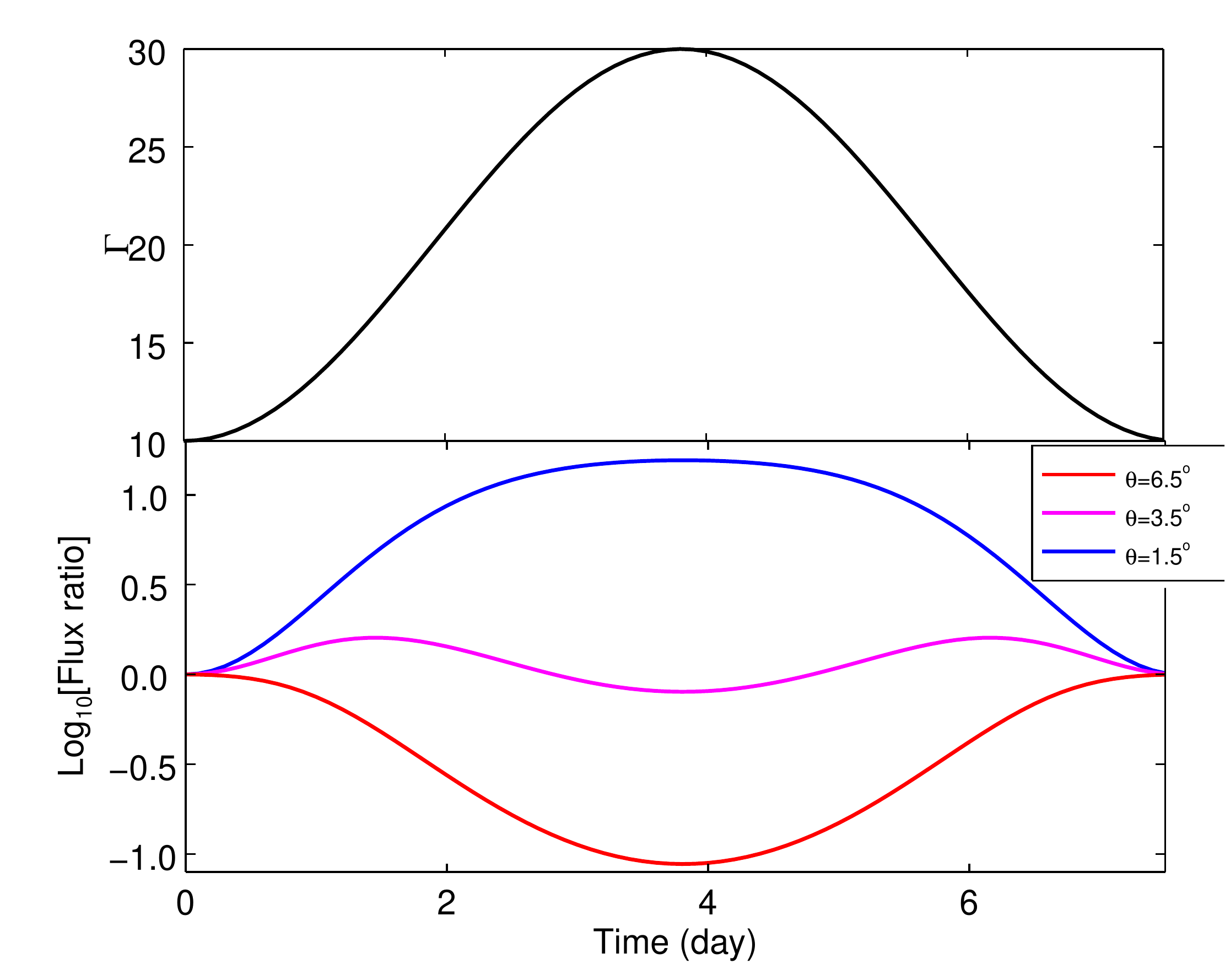}
 \caption{\textit{Left:} As the angle between the emission region and the line of sight decreases (top panel), the flux appears to flare as a result of relativistic beaming (bottom panel). The three curves correspond to the three different values of the bulk Lorentz factors. \textit{Right:} Similarly, flaring of flux (bottom panel) owing to the increase in the bulk Lorentz factor (top panel). The three curves correspond to the three different values of the angles of sight.}
\label{fig:beam}
\end{figure*}

The optical spectral slope is, in general, larger than 1 in LSPs, (in our case, OJ 49, S4 0954+658, TXS 1156+295, and 3C 279)  and less than 1 in case of HSPs (here, PG1553+113). For illustrative purposes, however, here the spectral index is taken as $\alpha \sim1$. If we assume the apparent flux rise happens purely due to the changes in the Doppler factor ($\delta$), which further can be related to the changes in the angle with the line of sight or/and the changes in the bulk Lorentz factor. Here as an illustration we treat two cases considering change in $\theta$ and $\Gamma$ separately.

\begin{itemize}
\item Change in the angle of the line of sight:\\
The angle between the emission region and the line of sight is allowed to gradually decrease, as approximated by $\rm{ \theta=\theta_{0}-Asin^{2}\omega t}$, where $ \theta_{0}$=5.7 and A=3.7 as shown in the top left panel of Figure \ref{fig:beam}. The resulting flux rise profiles for the three values of the bulk Lorentz factor, i.e. $\Gamma$=10, 15 and 20 represented by the blue, magenta and red curves, respectively, are shown in the lower left panel of Figure \ref{fig:beam}. Furthermore to approximate the flaring behavior that last about a week in the observer's frame, $\omega$ was chosen to be $15\times10^{-6}$ rad/sec.
 \item Change in $\Gamma$ :\\
 The bulk Lorentz factor of the dominant emission region is allowed to gradually increase, as approximated by $\rm{ \Gamma=\Gamma_{0}+Asin^{2}\omega t}$, with $ \Gamma_{0}$=10 and A=20, so that the plasma blob traveling at a speed of $ \Gamma=10$ accelerates to attain $ \Gamma=30$ and subsequently decelerates back to the previous speed. The evolution of $ \Gamma$ in time can be seen in the top right panel of Figure \ref{fig:beam}. As in the previous case, the value of $\omega$ was chosen to be $15\times10^{-6}$ rad/sec such that the event lasts about a week in the observer's frame. The resulting flux rise profiles for the three values of the angles of the line of sights, i.e. $\theta$=1.5, 3.5 and 6.5$^o$ are represented by the blue, magenta and red curves, respectively, shown in the lower right panel of Figure \ref{fig:beam}.
 \end{itemize}

\subsection{Stochastic flux variability and possible periodicity}
Blazars exhibit complex flux variability patterns with a possible mixture of several components such as general stochastic fluctuations, occasional large amplitude flares and possible quasi-periodic oscillations arising from various instabilities both in the disc and the jet. The general aperiodic and stochastic variability observed in multi-wavelength observations of AGN is largely represented by red-noise \citep[e.g. see][]{Isobe2015,Kelly2011,Bhatta2020}. The spectral power density of such variability is most consistent with a single power-law model, with negative spectral index ranging from $\sim$1--2. A negative power-law implies that the variability power grows towards longer timescales, meaning the source fluxes vary by larger amplitudes over yearly timescales compared to shorter ones, i.e. intraday, daily, weekly and monthly timescales. However it is interesting to note that in the particular case of the 3C 279 light curve in the bottom left panel of Figure \ref{Fig4}, the mean flux nearly remains stationary while the shorter term flux fluctuates rapidly. Similar observations can be made about PG 1153+113 when considering only the light curve before the vertical dashed line. It is only when we consider the full-length light curve that the longer term variability appears dominant. It is possible that the processes driving variability which can be characterized by a negative power-law index are distinct in origin compared to the commonly-observed variability characterized by positive power-law index. They could be signatures of the quasi-stationary isolated events which gets mixed with the general red-noise like variability.

Apart from the general aperiodic variability, Lomb-Scargle periodogram analysis of the light curves revealed hints of QPOs in some of the sources. However, the estimated significance of the peaks in the periodogram is moderate at about 90 \%. Nevertheless, it is important to note that in blazar light curves heavily dominated by red noise the actual QPO signals could appear relatively weak \citep[see discussion in][]{Bhatta2020}. The periodicity analysis indicates that in blazar 3C 279 the two periods, well above 90\% significance level, appear to be in a 3:2 ratio. In X-ray binaries such QPOs with periods in a 3:2 ratio are interpreted in terms of resonant oscillations of accretion flow \citep[e.g.][]{2005Ap&SS.300..143K}. Blazar QPOs on timescales of a few months are relatively rare, although yearly timescales QPOs have been reported in several blazars . Note that although QPOs in a timescales of a few days are more likely to originate at the accretion disc, we detect them through the jet emission. Possible interpretations of the blazar QPO are discussed in detail in our previous works \citep[see][and references therein]{Bhatta2020,Bhatta2019, Bhatta2017}.

\section{Conclusion}
\label{sec:6}
A variability study was carried out using optical observations of the blazars 3C 279, OJ 49, S4 0954+658, TXS 1156+295 and PG 1553+113 spanning several weeks. The light curves using observations acquired through our ground based telescope networks showed some of the extraordinary flaring events in which a flux change of nearly 10 times was observed within a timescale of $\sim$10 days. Flares in the sources OJ 49, S4 0954+658, and TXS 1156+295 were studied using a functional form of exponential rise and decay. Of these sources, slightly asymmetric flares of comparable normalized amplitude of $\sim 8$ were observed in the blazars OJ 49, S4 0954+658; whereas light curves of the source TXS 1156+295 revealed a fasted flux decay by $\sim 11$ normalized amplitudes within the timescale of four days. Such a rapid, large amplitude flux change reflects the most violent processes in the jet, e.g. shock waves and magnetic re-connection events. To explain the observed flares, qualitative descriptions of possible intrinsic and extrinsic scenarios were presented. In the source-intrinsic scenario, particle injection at the shock wave front can result in a flare in the source flux; whereas in the extrinsic scenario turbulent flow in the jet might lead to a change in the Doppler factor of a single, dominant energized cell such that a slight change in the speed or angle to the line of sight can yield a large observed change in flux. 

Furthermore, periodicity analysis was performed on the sources 3C 279, OJ 49, TXS 1156+295 and PG 1553+113 using Lomb-Scargle periodograms supplemented by a large number of simulated light curves generated using Monte Carlo methods. The result indicated that a strong spectral peak was detected at the characteristic timescale of $\sim$ 12 days to above 99\% significance level against power-law noise. In addition, two potentially significant periods 33 and 22 day, at a ratio of 3:2, were found to be detected at a significance level of $\sim 95\%$.

\section*{Acknowledgments}
We are grateful to the anonymous reviewer for their constructive comments, which helped improve the quality of the paper significantly. We acknowledge the support of the Polish National Science Centre through the grants UMO-2017/26/D/ST9/01178 (GB), 2018/09/B/ST9/02004 (SZ), and 2020/39 / B / ST9 / 01398 (DG). KM acknowledges JSPS KAKENHI grant number 19K03930.

\section*{Data availability}
The data used in this work can be shared on reasonable request to the
corresponding author.

%

\vspace{5mm}




\end{document}